\documentclass[%
 reprint,
 amsmath,amssymb,
 aps,
]{revtex4-2}

\usepackage{graphicx}% Include figure files
\usepackage{dcolumn}% Align table columns on decimal point
\usepackage{bm}% bold math
\begin{document}

\preprint{APS/123-QED}

\title{Burst dynamics, upscaling and dissipation of slow drainage in porous media}% Force line breaks with \\
\author{Knut J{\o}rgen M{\aa}l{\o}y$^{1,2}$}
\email{maloy@fys.uio.no}
\author{Marcel Moura$^{1}$}
\author{Alex Hansen$^{4}$} 
%\author{Tom Vincent-Dospital}
\author{Eirik Grude Flekkøy$^{1,3}$}
\author{Renaud Toussaint$^{5,1}$}

\affiliation{$^1$PoreLab, The Njord Centre, Department of Physics, University of Oslo, P.O. Box 1048 Blindern, N-0316 Oslo, Norway}
\affiliation{$^2$PoreLab, Department of Geoscience and Petroleum, Norwegian University of Science and Technology, 7031 Trondheim, Norway}
\affiliation{$^3$PoreLab, Department of Chemistry, Norwegian University of Science and Technology, 7031 Trondheim, Norway}
\affiliation{$^4$PoreLab, Department of Physics, Norwegian University of Science and Technology, 7031 Trondheim, Norway}
\affiliation{$^5$University of Strasbourg, CNRS, Institut Terre et Environnement de Strasbourg, UMR 7063, Strasbourg F-67084, France}

\date{\today}% It is always \today, today,

\begin{abstract}
We present a theoretical and experimental investigation of drainage in porous media. The study is limited to stabilized fluid fronts at moderate injection rates, but it takes into account capillary, viscous, and gravitational forces.   In the theoretical framework presented, the work applied on the system, the energy dissipation, the final saturation and the width of the stabilized fluid front can all be calculated if we know  the dimensionless fluctuation number, the wetting properties, the surface tension between the fluids, the fractal dimensions of the invading structure and its boundary, and the exponent describing  the divergence of the correlation length in percolation. Furthermore, our theoretical description explains how the Haines jumps' local activity and dissipation relate to dissipation on larger scales. 
%We also show that the energy budget is consistent between the work done on the system and the elastic energy released by out-of-equilibrium Haines jumps. 
\end{abstract}

\maketitle

%\tableofcontents
\section{Introduction}
Two-phase flow in porous media is crucial in a variety of sectors, ranging from fundamental research to applications in a wide array of industrial sectors such as fuel cell \cite{anderson2010} and solar cell technology \cite{hwang2012}, fiber-reinforced composite materials \cite{cantwell1991}, textile fabric characterization \cite{gibson1999}, prospection and exploration of oil and gas \cite{lake1989,afrapoli2011,yan2012,vasseur2013} etc. The description of flows inside natural porous media, such as soils and rocks, is also crucial for the study of groundwater flows \cite{guymon1994,bear1972} and the treatment of soil contaminants \cite{bear1987,jellali2001,nsir2012} but it also matters for everyday tasks such as making a cup of coffee \cite{gagne2021}. It is a multidisciplinary subject that has been investigated for decades by hydrologists, physicists, chemists, geoscientists, biologists, and engineers due to its practical importance and complexity. 
The structures observed are controlled  by the  forces involved, such as viscous \cite{chen1985,maloy1985,weitz1987,lenormand1988,lovoll2004,tallakstad2009,tallakstad2009A}, capillary \cite{lenormand1988,lenormand1985,furuberg1988,lovoll2004,tallakstad2009,tallakstad2009A,moura2017A,xiao2021}, and gravitational forces \cite{wilkinson1984,birovljev91,frette1992,wagner1997,auradou1999,muharrik2018}, as well as wetting properties \cite{zhao2016,cottin2011,holtzman2015,holtzman2020,gennes2004,primkulov2018,primkulov2019} and changes in the local geometry of the porous medium \cite{rabbani2018,lu2019}. 
The structures vary in shape  and complexity \cite{lenormand1988,lenormand1989,zhao2016,maloy1985,chen1985,payatakes1980,sandnes2007,sandnes2011,odier2017,primkulov2018,ferer1993,ferer2003}, from compact to ramified and fractal \cite{mandelbrot1982,feder1988}. The fractal nature of porous media is itself important for a number of applications \cite{yu2008,xiao2019}, such as electrolyte diffusion through charged media \cite{liang2019} a topic of relevance for the development of modern battery technology \cite{armand2020}.

In most practical applications of porous medium physics, the typical length scales where our interest lies is substantially larger than the scale where the relevant physics is taking place. Oil reservoirs or water aquifers are in the range of kilometers while the typical pore sizes are commonly in the micrometer scale, about nine orders of magnitude smaller. How do we deduce the flow behavior at large scales from small-scale physics?  The usual way  of solving these problems is  a top-down approach using  Darcy´s law \cite{darcy1856}  on a mesoscopic level. However this  approach  does not take into account local fluctuations, like capillary or viscous fluctuations, which are averaged out. 
In this manuscript we take an alternative bottom-up approach where we  emphasize on the pore-level capillary fluctuations and  compare those  fluctuations  with the  characteristic forces  which are set up by the external fields on the whole system. Examples of such forces are gravitational or  viscous fields. We will also limit our discussion to the stable drainage regime, which occurs when a nonwetting fluid displaces a wetting fluid, and when the viscous and/or gravitational forces stabilize the displacement front between the two fluids. This approach was  introduced in the late 1970s  \cite{gennes1978,chandler1982} and its theoretical development benefited greatly from invasion percolation models \citep{wilkinson1983}. On this subject, other relevant experimental, numerical, and theoretical articles have since been published \cite{wilkinson1984,birovljev91,frette97,aker2000C,aker2000B,meheust2002}. The dimensionless fluctuation number $F$, introduced in \cite{auradou1999,meheust2002},  quantifies the ratio between the viscous and/or gravitational field and the capillary pressure fluctuations.  The characteristic length  scale $\eta$ which describes the width of the invasion front, and which depends on $F$,  is of central importance to calculate  the saturation behind  the front, which in turn gives a measure of the sweeping efficiency of a given drainage process, a quantity of great interest in a number of applications.  The local structures will be  fractal on length scales  smaller than $\eta$ with a crossover to  a  homogeneous behaviour on length scales larger than $\eta$.  Knowing the scaling of the fluid structures up to the length scale $\eta$ allows one to have full control of the energy balance of the problem and calculate large-scale quantities such as the final saturation behind the front, the  dissipation or the entropy production, and the work required to move the front forward.   

The dynamics of slow displacement in porous media has been observed to occur in an intermittent  manner by so called Haines jumps \cite{haines1930,maloy92,aker2000C,furuberg96,moebius2014,moebius2014a,moebius2012,berg2013,bultreys2015,zacharoudiou2018,moura2017,berg2020}. 
The invasion percolation model \cite{gennes1978,chandler1982,wilkinson1983}  describes well the structure of slow displacement in porous media \cite{lenormand1985}, but it does not describe the dynamics realistically because the invasion is limited to one pore at a time. However it is possible to introduce a  realistic interpretation of time in a modified invasion percolation model  \cite{maloy92} by introducing a  constant $\kappa$ relating the  volume change at the interface   with a corresponding change in pressure.  
%We found both in simulations and experiments \cite{maloy92} \cite{furuberg96} that the  pressure fluctuations of the Haines jumps  follow a power law distribution with an exponential cutoff function.  In this manuscript we have used this  distribution functions of the pressure fluctuations together with the known fractal dimensions of the invasion structure, and the fractal dimension of the contours of the trapped clusters to calculate the work, and the dissipation  when the displacement front has reached a steady state.  
Ref.~\cite{maloy92,furuberg96} found both in simulations and experiments that the  pressure fluctuations of the Haines jumps  follow a power law distribution with an exponential cutoff function. 

%In the present work we use a new approach where  the elastic energy (surface energy) released by the invasion front and the surface energy created by the new pores invaded in the Haines jump can be used to calculate the energy dissipation of a single jump.
In the present work we use a new approach to calculate the dissipation and the energy of the surfaces left behind the front  by  considering the elastic  energy  (surface  energy)  released  by  the  Haines Jumps when the front is in a steady state regime. In between jumps all the external work goes into building up the elastic (surface) energy of the fluid front.  When the invading fluid front is in a steady state, its average length is constant, and the work applied to the system (for example by an external pump) must be equal to the elastic energy released by the invasion front (green line in Fig.~\ref{syst}).  As a result, using the distribution function of the capillary pressure fluctuations and performing a bottom-up approach to integrate up the elastic energy released by  the bursts, we can check the consistency of our theory as well as the distribution of dissipative events. We found as expected  that the total elastic energy released by the bursts is equal to the work  $W= \langle p \rangle \Delta V$, where $\langle p \rangle$ is the average pressure across the model and $\Delta V$ is the volume change of the invading fluid corresponding to the interface motion. This result provides an important consistency check for the analysis. The dissipation can then be calculated by subtracting the generated  surface energy  from the total applied work. 
The surface energy is directly computed by the scaling of the invasion structure, which is given by the invasion structure's fractal dimension, also measured in our work.  As a result of the theory and quasi-two-dimensional experiments, we discovered that the work $W$, dissipation $\Phi$, and the energy of the surfaces left behind the front $E_s$ are all proportional to the number of pores $S$ invaded. Up to the characteristic length scale $\eta$, the spatial scaling of $S$ is fractal and for larger length scales the scaling becomes proportional to the system size instead. We then show that we can explicitly calculate $W$, $E_s$, and $\Phi$ if we know the surface tension of the fluids and the wetting properties from the experiments. We further discovered an important analytical result: that the ratios $E_s/W$ and $\Phi/W$ are both independent of system size.

\section{\label{sec:level1} Experimental Technique}
The majority of the experimental results presented in this paper are taken from data published in the past in our group. 
The system is composed of a modified Hele-Shaw cell \cite{heleshaw1898}, filled with a monolayer of glass beads with diameters in the range $[1.0 mm, 1.2 mm]$. 
The glass beads are randomly distributed in the cell gap and the voids in between the beads form the porous network. This quasi-two-dimensional geometry allows for the direct visualization of the fluid phases, by means of regular optical imaging using a digital camera. For about the model construction, see for instance Refs.~ \cite{lovoll2004} and \cite{moura2015}.

\begin{figure}
\centering
\includegraphics[width=1.0\linewidth]{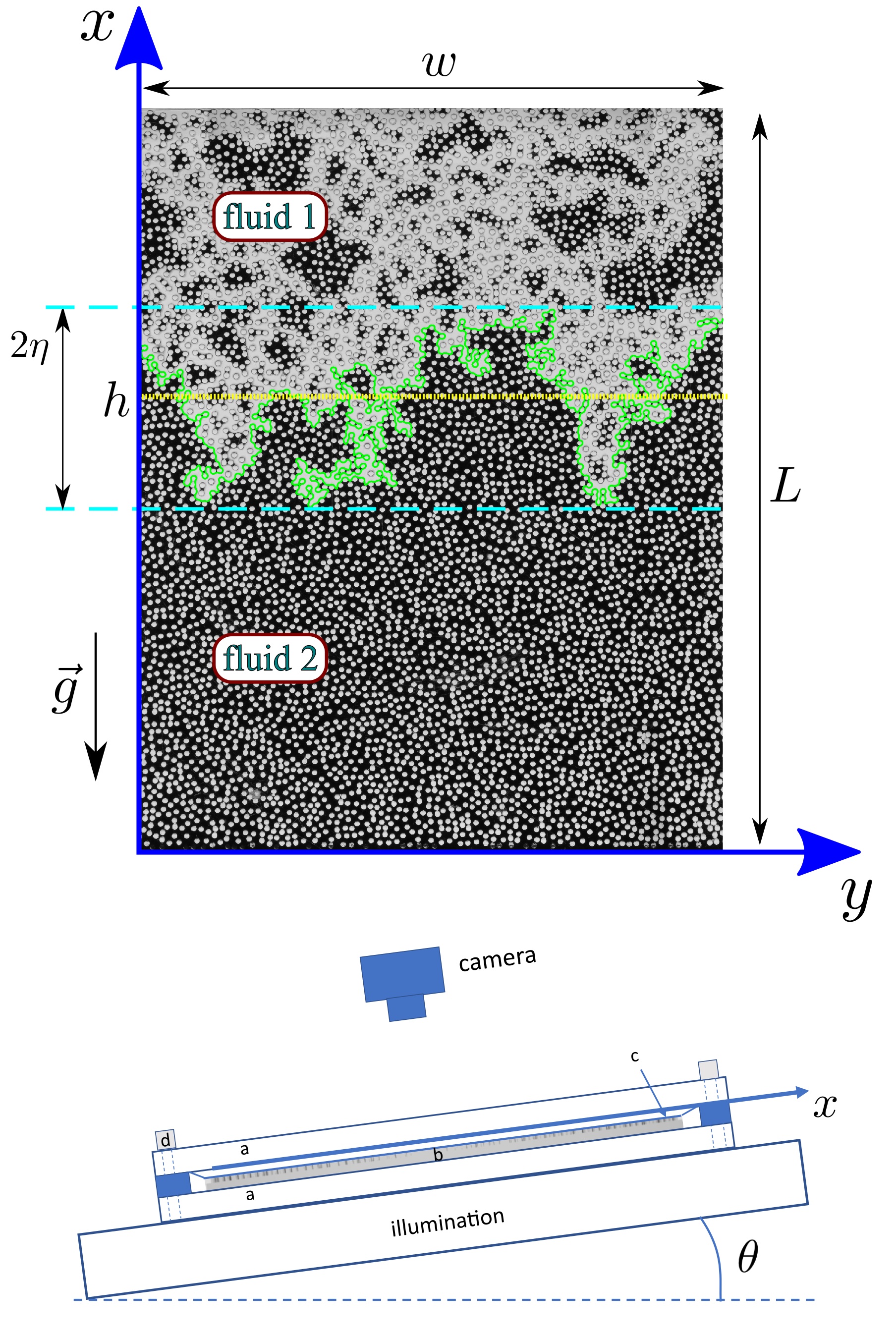}
\caption{Upper figure: A nonwetting fluid 1  invades another wetting fluid 2 in a porous model  of width $w$, length $L$ and coordinate system (x,y). A syringe pump is connected to the lower side and the model is open to air on the upper side. The model can be tilted with  an effective gravitational constant $g=g_0 \sin(\theta)$ along the x direction of the model, where  $g_0=9.82 m/s^2$.  The average position of  the front is $h$ and the  width of the front is  $2 \eta$. Lower figure: Cross section of the experimental model system a) shows two transparent Plexiglass plates, b) the 3D printed porous model, c) a PVC film kept under pressure to close the model and d)  screws that clamp the model. The model is illuminated from below and pictures are taken from above.}
\label{syst} 
\end{figure}

In this paper we also present results produced using an alternative stereolithography 3D printing technique to create the porous structures. We have employed a Formlabs Form 3 printer to produce models in a transparent plastic material (Clear Resin FLGPCL04). This technique allows us to control the geometry of the porous network and in particular to tune its porosity.  In the experiments presented here we have made quasi-two-dimensional models where cylinders are distributed with a Random Sequential Adsorption algorithm \cite{hinrichsen1986} where we can set the minimum distance between the cylinders (see Fig.~\ref{syst}), typically chosen as $0.3mm$. The cylinders height and diameter were both chosen to be $1mm$. The spatial resolution of these models is about $0.09mm$ and the models are constructed to optimize the visualization of the pores, which are seen from a top-down view. The 3D printed porous model is placed between two thick Plexiglass plates which are clamped around the edges using screws to give robustness to the setup and ensure the quasi-two-dimensional geometry. A flexible PVC film is placed between the porous model and the top Plexiglass plate. This film plays an important role: due to its flexibility, when the screws around the model are fastened the PVC film gets in contact with the top of all cylinders, thus ensuring the appropriate sealing of the model. The film has similar wetting properties as the 3D printing material used in the construction of the model. Fig.~\ref{syst} shows a typical snapshot of an experiment and a diagram of the setup. We also define in the upper part of the figure the model's width $w$, length $L$, the invading front (green line), its average position $h$ and width $2\eta$.

The porous network is initially fully saturated with a wetting viscous liquid composed of a mixture of glycerol ($80\%$ in weight) and water ($20\%$ in weight). The kinematic viscosity, density and surface tension (with respect to air) are $\nu = 4.25\cdot10^{-5}\, {m^2/s}$, $\rho = 1.205\, {g/cm^3}$ and $\gamma = 0.064 \,{N/m}$. The wetting liquid is dyed with a dark blue colorant (Nigrosin), to aid visualization. Air is used as the nonwetting phase. The contact angle measured inside the wetting phase is $\psi=70^{\circ}$. During an experiment, the liquid phase is withdrawn with a syringe pump (Harvard Apparatus) at a constant flow rate, leaving the model from a width-spanning channel at the bottom end of the cell. Air enters the model from the top, through another width-spanning channel that is open to the atmosphere.

\section{\label{sec:level2} Surface energy, dissipation and burst dynamics}
\begin{figure}
\centering
\includegraphics[width=1.0\linewidth]{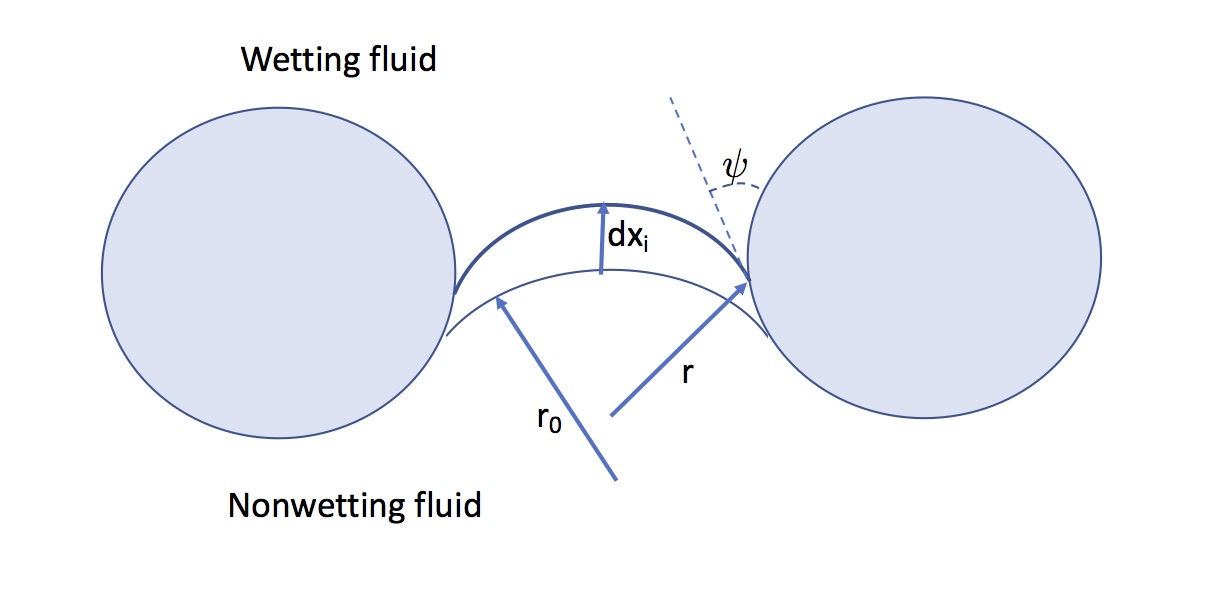}
\caption{The figure illustrates a nonwetting fluid invading a wetting fluid at a single  pore-throat at different radius of curvature $r_0$ and $r=r_0-dr$.
The distance moved between the two front positions is $dx_i$ and $\psi$ is the wetting angle}
\label{geomet} 
\end{figure}

Consider a single pore identified by the index $i$ in which a nonwetting fluid displaces a wetting fluid as illustrated in Fig.~\ref{geomet}. The front moves from a position with a radius of curvature  $r_0$ to one with a radius of curvature  $r=r_0-dr$.  This results in a  volume change  $dv_i$ of the nonwetting fluid, and a corresponding volume change $-dv_i$ of the wetting fluid in that pore. 
Assume that we measure the distances in the normal (radial) direction from each of the points on the front defined with radius $r_0$ to the front defined by the radius of curvature $r=r_0-dr$. The longest of these distances we denote as $dx_i$.  The distance $dx_i$ to lowest order in $dr$ is given by
\begin{equation}
    dx_i= \left. \frac{dx_i}{dr} \right|_{r_0}dr= \alpha_i dr \; ,
\label{dx}
\end{equation}
where 
\begin{equation}
\alpha_i(\psi,p_0) =\left. \frac{dx_i}{dr} \right|_{r_0} \; 
\end{equation}
depends on the wetting angle $\psi$ (See Fig.~\ref{geomet}) and the capillary pressure $p_0$ given by the  Young-Laplace equation $p=\gamma(1/r+1/r_1)$ \citep{bear1972} for $r=r_0$. Here $r$ is the in-plane,   and $r_1$ the out of plane radius of curvature assumed to be constant in the quasi-two-dimensional experiments.  

We can then  calculate the corresponding increase in capillary pressure
\begin{equation}
    dp=\frac{\gamma dr}{r_0^2} \;.
    \label{kelvin}
\end{equation}
Then by using  Eqs. (\ref{dx}) and  (\ref{kelvin}) we get
\begin{equation}
    dp = \frac{\gamma}{r_o^2 \alpha_i}   dx_i \; .
\end{equation}
 The  volume change $dv_i$ of the invading fluid in the pore can be written as 
\begin{equation}
    dv_i= {\cal A}_i(\psi,p_0)  dx_i , 
    \label{apx}
\end{equation}
where ${\cal A}_i(\psi,p_0)$ is the surface area in pore $i$ which is function of the wetting angle and the capillary pressure $p_0$.  We then get 
\begin{equation}
    dp = \frac{\gamma}{r_o^2 {\cal A}_i \alpha_i}  dv_i \; ,
\end{equation}
such that
\begin{equation}
    d v_i =  \kappa_i dp \; ,
\label{kappapv}
\end{equation}
where the capacitive volume
\begin{equation}
\kappa_i(\psi,p_0)=\frac{r_0^2 {\cal A}_i(\psi,p_0) \alpha_i(\psi,p_0)}{\gamma} \; , 
\end{equation}
gives the fluid volumetric change per unit capillary pressure in  pore $i$ \citep{maloy92}.
\begin{figure}
\centering
\includegraphics[width=1\linewidth]{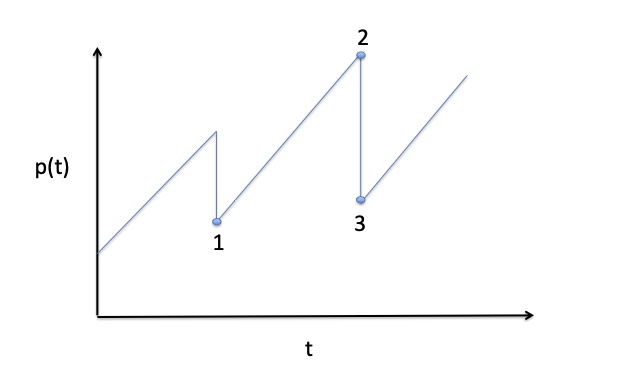}
\caption{The figure illustrates the dependence of the pressure across the model $p(t)$ and the time $t$. The front is  in equilibrium between point 1 and point 2 and reaches the threshold value of one of the pores along the front at point 2.  A burst will appear between point 2 and 3 and the front is there out of equilibrium.}
\label{sample} 
\end{figure}

Let $U_i(x)$ be the elastic energy (surface energy)  of the interface in pore $i$, and  $dx_i$   a small displacement of the interface between the two fluids  due to an external pump driving the system at a low flow rate such that viscous forces can be neglected. Assume that the system is in mechanical equilibrium. The work performed by the pump will then increase the elastic energy due to this displacement.
%We will assume that the flow rate $q$ is so small that we can neglect the viscous forces.  
The elastic energy associated to pore $i$ having interface located at the position $x_i$  would then be
\begin{equation}
    U_i(x_i)=U_i(0)+\int_0^{x_i} p {\cal A}_i  dx_i \; ,
    \label{elasticU}
\end{equation}
where $p$ is the capillary pressure across the interface, and $U_i(0)$  is the elastic energy when $x_i=0$. 
Using Eqs.  (\ref{kappapv}), (\ref{elasticU}) and (\ref{apx}) we get
\begin{equation}
  U_i(x_i)=U_i(0)+\kappa_i \int_{p_0}^{p} p^* dp^*=U_i(0)+\frac{\kappa_i}{2} p^2 \; .
\end{equation}
We have assumed $p_0=0$ without loss of generality since $\kappa_i p_0^2/2$ also  can be included in $U_i(0)$.  If instead of considering a single pore we consider the work done on a front having $n$ pores, the total elastic energy  $U_t$ will be the sum of the contributions from all pores belonging to the invasion front (green line in Fig.~\ref{syst}).  We then get
\begin{equation}
    U_t=U_t(0)+ n \frac{\kappa}{2} p^2 \; ,
\end{equation}
where 
\begin{equation}
    \kappa=(\sum_{i=1}^{n} \kappa_i)/n  \; ,
\end{equation}
where  $\kappa_i$ is the capacitive volume of pore $i$ so that $\kappa$ is the average capacitive volume over the $n$ interface pores.  
The capillary pressure $p_2$ at the time $t_2$ and the capillary pressure $p_1$  at the time $t_1$ are  related by
\begin{equation}
    p_2-p_1=\frac{q(t_2-t_1)}{n \kappa} \;, 
\end{equation}
where $q$ is the imposed volumetric flow rate. 
Since we have assumed the system to be in  mechanical equilibrium,  the interface will slowly increase its capillary pressure without any bursts (Haines jumps) \cite{haines1930}.  The capillary pressure  will then build up from $p_1$  at $t_1$ to $p_2$ at $t_2$ due to the work $W$ performed by the external pump, see Fig.~\ref{sample}. This work will then increase the total elastic energy (surface energy)  of the interface from $U_t(t_1)$ to $U_t(t_2)$

\begin{multline}
    U_t(t_2)-U_t(t_1)=W=n \kappa \int_{p_1}^{p_2} p^* dp^*=\\ 
    \frac{n \kappa}{2}(p_2^2-p_1^2) \; .
\end{multline}

After some time the interface will however  reach a situation  where the capillary pressure in one of the pore-throats is at the threshold value $p_t$. The interface at that pore will then become unstable, and the invading fluid will move into one or more neighboring pores.
At this time, the capillary pressure at the interface of the growing burst will be lower than the capillary pressure at the other parts of the interface. This produces a local velocity field that extends from the growing burst to the other parts  of the interface. The interface will then back-contract, beginning with the pores closest to the pore where the burst begins and spreading out through the interface until it reaches equilibrium.  Then the capillary pressure is again the same at all pores along the interface. Let us assume that the burst starts at  time $t_2$ at a capillary pressure $p_2=p_t$ and that, after the burst, the system reaches another equilibrium state at capillary pressure $p_3$. During the burst the elastic energy of the part of the fluid interface which is back-contracting will be reduced. This energy reduction is equal to
\begin{multline}
    U_t(p_2)-U_t(p_3)=\frac{\kappa n(t_2)}{2} (p_2^2-p_3^2)= \\ \kappa n(t_2) (p_2-p_3) (p_2+p_3)/2 \; ,
\label{enred}
\end{multline}
and will go to  creation of the surface energy of the new burst in addition to  viscous dissipation due to the flow that takes place during the burst.  Here $\Delta p=p_2-p_3$ is the pressure drop in the burst   and $(p_2+p_3)/2$ is the average pressure during the burst.

\begin{figure}
\centering
\includegraphics[width=1\linewidth]{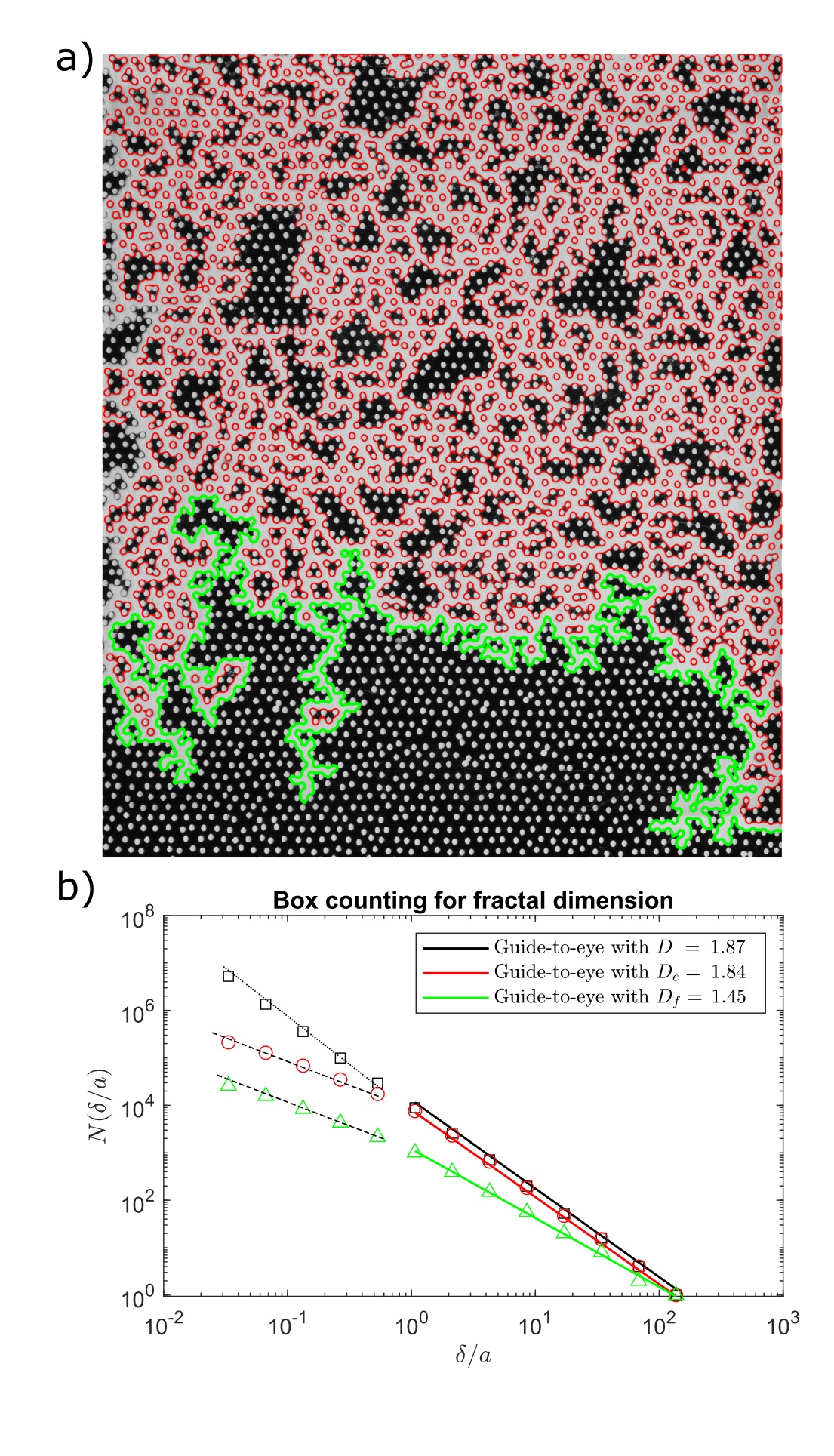}
\caption{a) Tracking of the invasion front (green) and of the in-plane invasion contours left behind the invasion front (red). Notice that some isolated beads are included in the tracking. b) Measurement of mass fractal dimension $D$, the fractal dimension $D_e$ of the contours in a) and the fractal dimension $D_f$ of the main invasion front (green line). The dotted and dashed lines have slopes of -2 and -1 respectively.}
\label{fig:fract_De} 
\end{figure}

The surface energy of one single burst in our quasi-two-dimensional experiments has two contributions,  one from the  area of the burst $2 \gamma_2 a^2s$ and one from the interface contour  $ a \gamma_1 b s_e$. Here $a$ is the average in-plane pore size, $b$ is the height of the cylinders (or the  diameter of the beads in the case of the previously published experiments based on a monolayer of glass beads, see Sec.~\ref{sec:level1}), $s$ is the number of pores in the burst considered, and $as_e$ is the length of the invasion contour of the burst (including trapped clusters). The factor 2 in the term $2 \gamma_2 a^2s$ is due to the creation of two new solid-air interfaces after the burst, one at the bottom and one at the top. The need for two different surface energies values, $\gamma_1$ and $\gamma_2$, is explained below. 

In Fig.~\ref{fig:fract_De} a) we show  the tracking of the invasion front (green line), and the in-plane invasion contour left behind the front (red line).  The fractal dimensions $D$ of the invading fluid phase (air), $D_e$ of the invasion contours, and $D_f$ of the invasion front are shown in Fig.~\ref{fig:fract_De} b), where a box counting technique was employed in the measurement \citep{feder1988}. We have obtained the values $D=1.87 \pm 0.10$, $D_e=1.84 \pm 0.10$ and $D_f=1.45 \pm 0.10$. The dotted and dashed lines on the left correspond to the exponents -2 and -1 respectively. These are expected as the fractal nature of the invasion structure has a lower bound at the pore-size $a$. For boxes of size $\delta<a$, we recover the intrinsic 2 dimensional nature for the invading structure mass (dotted line) and the 1 dimensional nature of the perimeters of the front and internal contours (dashed lines). 

\begin{figure}
\centering
\includegraphics[width=1\linewidth]{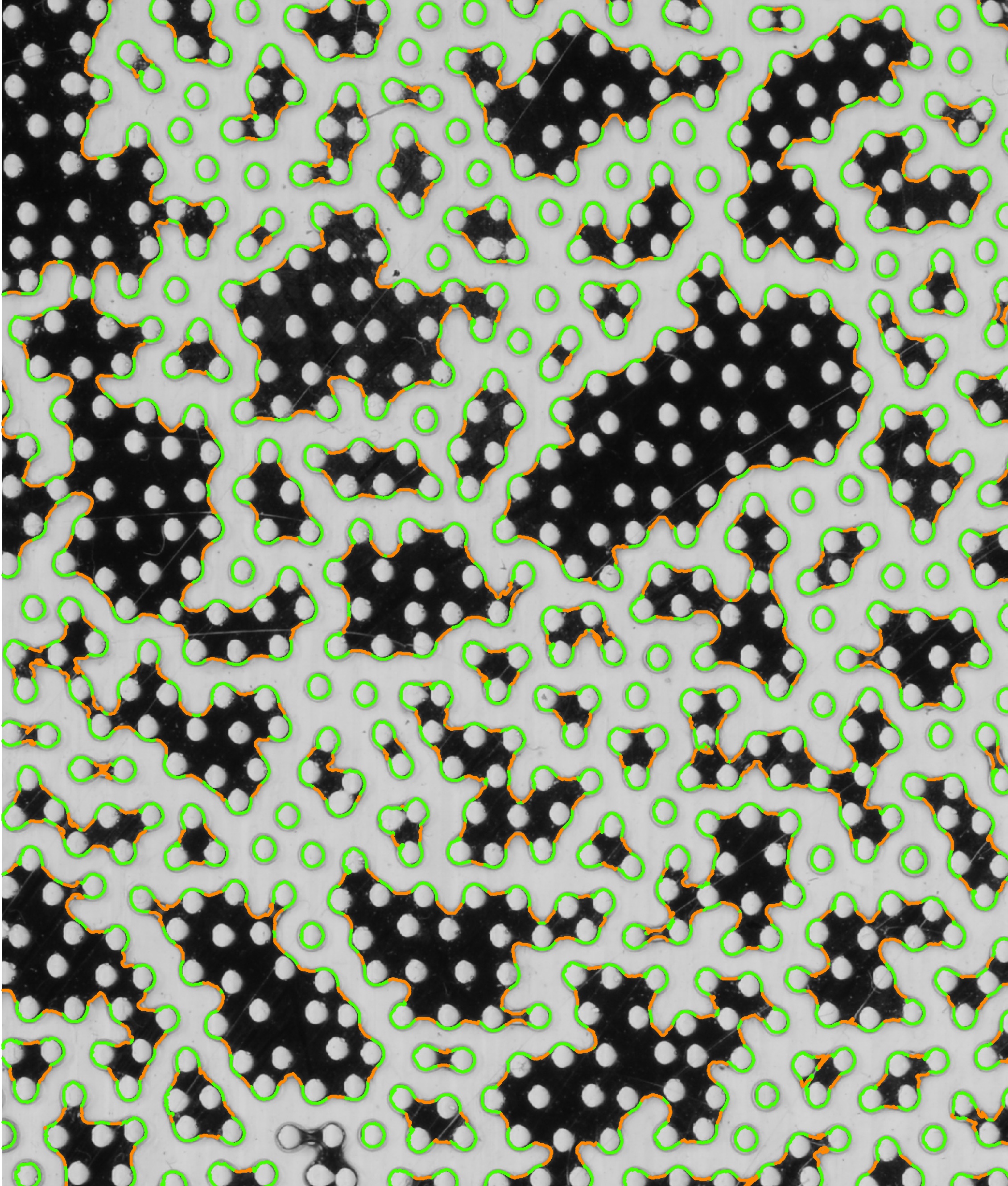}
\caption{Tracking of the internal interfacial contours between the nonwetting and wetting phases (orange) and the nonwetting and solid phases (green).}
\label{fig:greenorange} 
\end{figure}

When a fluid film is left behind  the invasion front the surface tension $\gamma_1=\gamma_2=\gamma$ which is the surface tension between the two fluids.  However  when no film is  left behind the invasion front, $\gamma_2=\gamma_{ns}-\gamma_{ws}$, where $\gamma_{ns}$ is the interface tension between the nonwetting fluid and the solid and  $\gamma_{ws}$ is the interface tension between the wetting fluid and the solid. Using the Young equation \citep{bear1972} we can also write the previous relation as  $\gamma_2 = \gamma \cos(\psi)$, where $\psi$ is the static contact angle at the liquid-solid-air triple line.  In the situation in which no wetting fluid film is left behind the invasion, the in-plane interface contour (red lines in Fig.~\ref{fig:fract_De}) will be partly formed by nonwetting-wetting segments and partly formed by nonwetting-solid phase segments. This division is exemplified in Fig.~\ref{fig:greenorange}, where we have split the contours shown in red in Fig.~\ref{fig:fract_De} into orange and green segments, where orange denotes the interface between the nonwetting and wetting phases and green the interface between nonwetting and solid phases. Let us define the ratio $\epsilon=A_{ns}/A_t$, where $A_{ns}$ is the interface's surface area between the nonwetting phase and the solid phase, and  $A_t$ is the total surface area of the interface. In Fig.~\ref{fig:greenorange}, $\epsilon$ corresponds to the ratio between the total length of green lines to the total length of both green and orange lines added. We then get an effective surface tension $\gamma_1=(1-\epsilon)\gamma +\epsilon(\gamma_{ns}-\gamma_{ws})$ for the total internal interfacial contours. Using again the Young equation \citep{bear1972} we can rewrite this expression as  $\gamma_1=(1-\epsilon)\gamma +\epsilon\gamma\cos(\psi)$.

We will in the following consider a situation where we have a long channel containing the porous medium with the invading fluid entering from one of the short sides of this channel and with the outlet on the other side. We will further assume that we follow the displacement front. The front  will have a width of the order of the width $w$ of the channel. Apart from an initial transient  we can then consider the front to be in a steady state regime with a constant average length $n$ (measured in number of pores).  Since the elastic energy of the  growing interface on average must be constant, the work $W$ applied by the pump must equal the  elastic energy released by the bursts which is equal to the  dissipation $\phi$ plus the surface energy needed to create the the new interface $E_s$, such that
\begin{equation}
    W=\Phi+E_s \; .
    \label{ebudget}
\end{equation}
It is important to note that $W$, $\Phi$, and $E_s$ are average quantities that do not account for fluctuations, and that $\phi$ and $E_s$ are averaged over a large number of bursts. 

From \cite{maloy92} we know that the distribution function of the burst sizes $G(s)$ is directly linked to the distribution of the capillary pressure drops ${\cal F}(\Delta p)$ during the bursts since 
\begin{equation}
s=n \kappa \Delta p /\bar{v}  \; ,
\label{bup}
\end{equation}
where $\bar{v}$ is the average single pore volume averaged over  the invading structure,  and 
\begin{equation}
    {\cal F}(\Delta p)=G(s) \frac{ds}{d \Delta p} \; .
    \label{bdist}
\end{equation}

\begin{figure}
\centering
\includegraphics[width=1\linewidth]{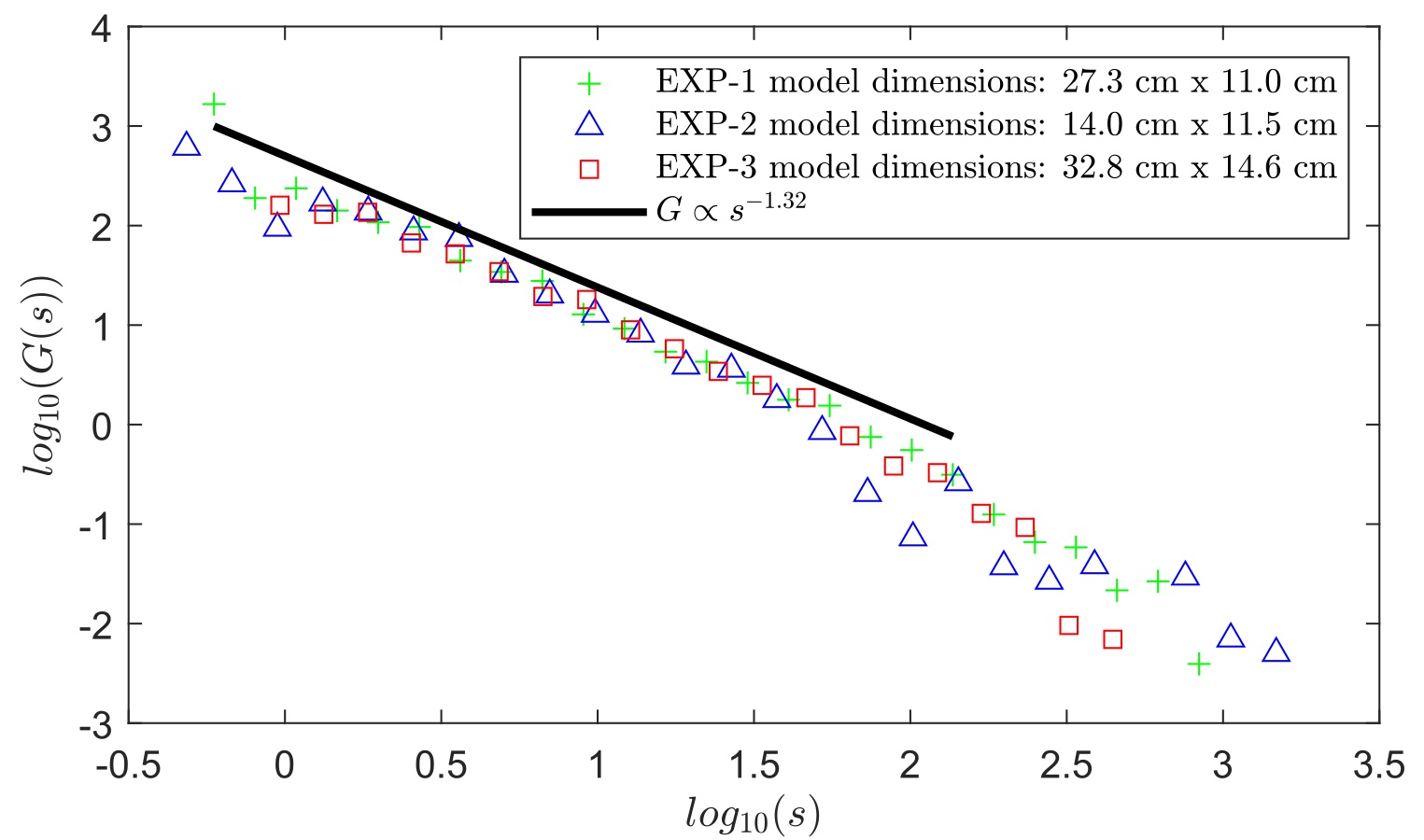}
\caption{Burst size distribution $G(s)$ from slow drainage experiments.  The line  has a slope of $-1.32$ corresponding to the results predicted by the analytical prediction  Eq.~(\ref{tau}). The data for EXP-2 (blue triangles) has been shifted vertically by 0.4. Data from Ref.~\cite{moura2017}.}
\label{Marcel-burst} 
\end{figure}

We found \cite{aker2000C,maloy92,furuberg96,moura2017}, both in  simulations and experiments  that the burst size distribution $G(s)$ is given by the scaling relation
\begin{equation}
    G(s)=s^{-\tau} H(\frac{s}{s^*})  \; ,
\label{gfrel}
\end{equation}
where the cutoff size of the bursts $s^*$ is \cite{maloy92}
\begin{equation}
    s^*=n \kappa \Delta p^*/\bar{v}= \left( \frac{n \kappa}{\bar{v} N(p_c)}\right) ^{\frac{\nu D}{1+ \nu D}} \; .
\label{scut}
\end{equation}
where $N(p_c)$ is the value of the normalized capillary pressure threshold distribution taken at the critical percolation threshold pressure $p_c$, and $\nu=4/3$ is the exponent describing the divergence of the correlation length in percolation \citep{stauffer1994}. 
Since the capillary pressure threshold distribution $N(p_t)$ is normalized, the width of the  distribution $\sigma \approx 1/N(p_c)$. 
Therefore the cutoff size of the bursts will increase with $\sigma$ according to Eq.~(\ref{scut}) as $s^* \propto \sigma^{0.78}$. 
Martys et al. \cite{martys1991} derived the analytical form 
\begin{equation}
    \tau=1+ \frac{D_f-1/\nu}{D} \; , 
    \label{tau}
\end{equation}
where $D$ and $D_f$ are, respectively, the fractal dimensions of the growing cluster and its front (seen as the green line Fig.~\ref{syst}). Using the literature values $D=1.82$ \cite{lenormand1985,wilkinson1983} and  $D_f=4/3$ \cite{furuberg1988,birovljev91,frette97},  we obtain $\tau=1.32$, consistent with our experimental data shown in Fig.~\ref{Marcel-burst}.  

\begin{figure}
\centering
\includegraphics[width=1\linewidth]{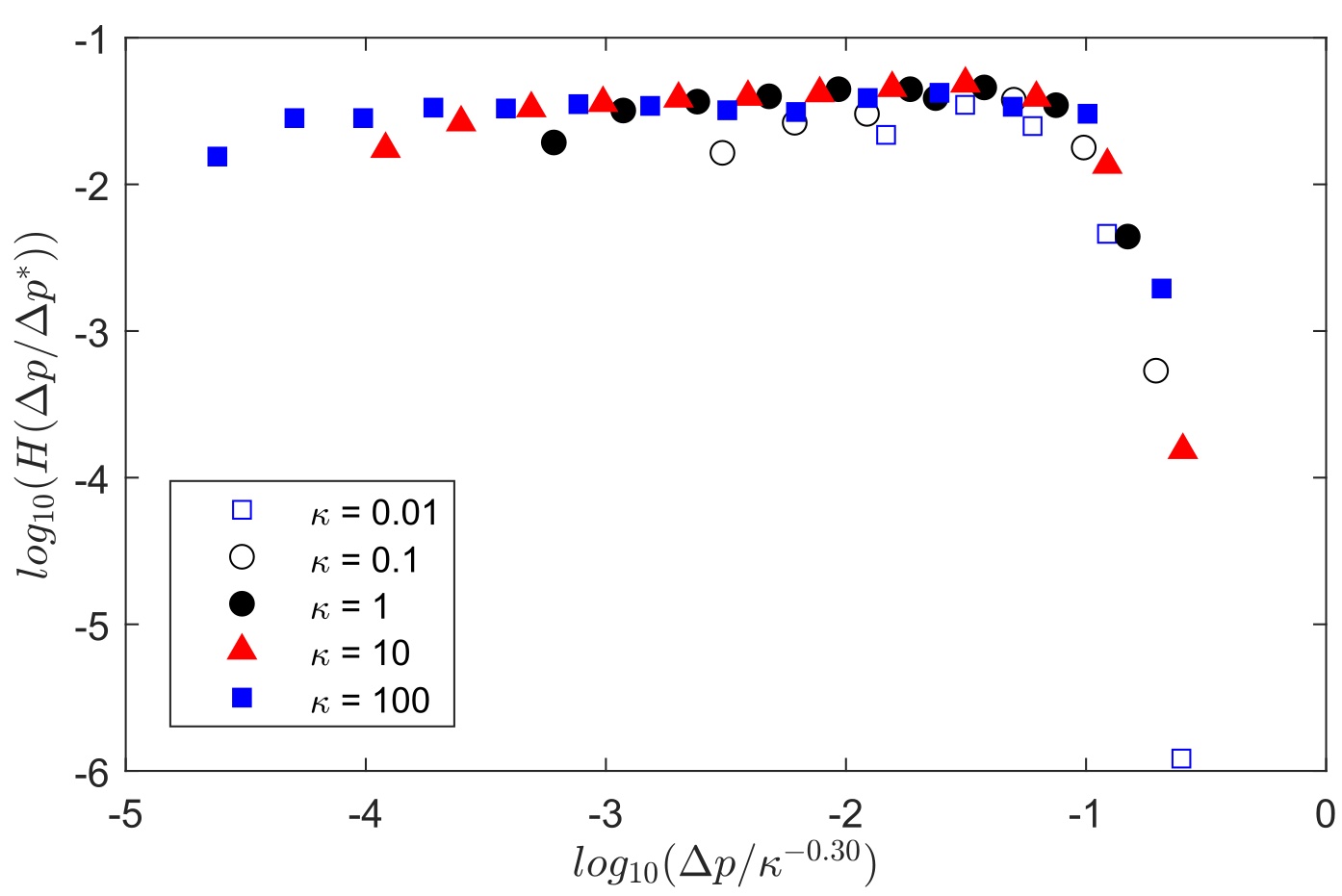}
\caption{The dependence of the crossover function $H(\Delta p/\Delta p^*)$ of the pressure jump distribution in a modified invasion percolation simulation \cite{maloy92}.  For each value of $\kappa$, the points represent averages over five independent simulations on $200 \times 1500$ lattices. The scaling exponent $0.30 \approx 1/(1+\nu D)$.
Data is taken from \cite{maloy92}. }
\label{Furu} 
\end{figure}

We therefore get from  Eqs. (\ref{bup}), (\ref{bdist}), (\ref{gfrel}) and (\ref{scut})  
\begin{equation}
    {\cal F}(\Delta p)= \Delta p^{-\tau} H(\frac{\Delta p}{\Delta p^*}) \left(\frac{n \kappa}{\bar{v}} \right) ^{(1-\tau)} \; , 
    \label{fdist}
\end{equation}
where Eqs. (\ref{scut}) and (\ref{bup}) give the cutoff pressure $\Delta p^*$
\begin{equation}
    \Delta p^*= \frac{1}{N(p_c)}\left( \frac{n \kappa}{N(p_c) \bar{v}}\right) ^{\frac{-1}{1+\nu D}} \;. 
    \label{pstar}
\end{equation}
Fig.~\ref{Furu} shows the scaling function $H(\Delta p/\Delta p^*)$ for various values of $\kappa$ from  invasion percolation simulations \cite{maloy1985} which confirms (\ref{fdist}) and (\ref{pstar}).

\begin{figure}
\centering
\includegraphics[width=1\linewidth]{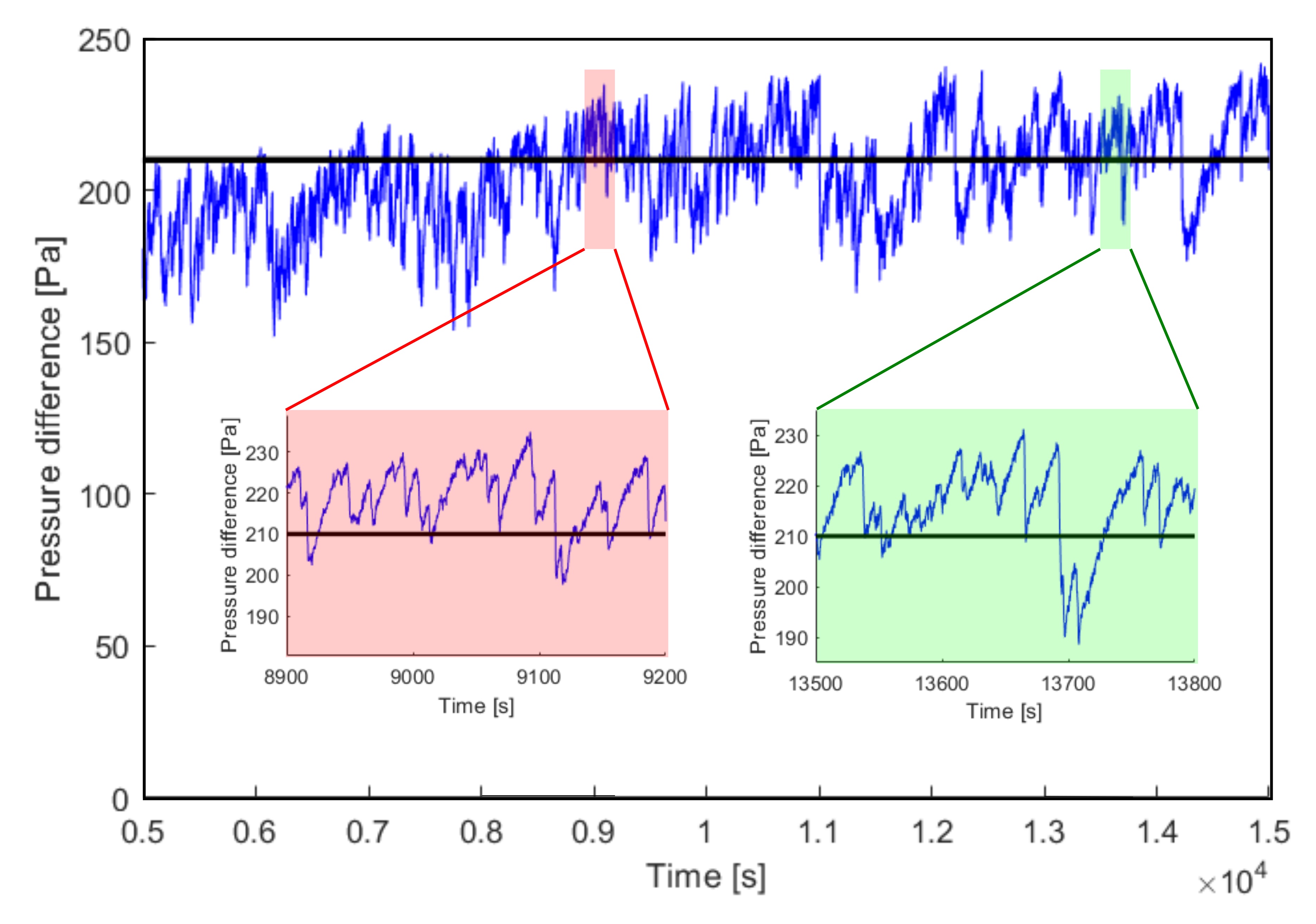}
\caption{Typical evolution of the pressure signal during an experiment. In this specific case, the liquid was withdrawn from the bottom at a rate of $1ml/h$ and the model was positioned horizontally, i.e.,  $\theta=0^{\circ}$. The thick horizontal line denotes a characteristic average pressure during the bursts $\langle p \rangle=210 Pa$. The left and right insets (shaded respectively in red and green) show zoomed in sections of the signal, where we can see the characteristic Haines jumps. The signal looks statistically similar in both insets, thus justifying the assumptions made in Eq.~(\ref{work2}).}
\label{pressure_signal} 
\end{figure}

We will in the following use ${\cal F}(\Delta p)$ to calculate the work.   Assume that the viscosity is low such that  the total time of the bursts is short, and can be neglected compared to the total time considered. 
The work can then be calculated by averaging Eq.~(\ref{enred})
\begin{equation}
    W=M n \kappa \langle p \rangle  \int_0^{\infty} \Delta p {\cal F}(\Delta p) d\Delta p \; ,
    \label{work2}
\end{equation}
where $M$ is the total number of bursts during the time considered.  Here we have assumed that the  average pressure in each burst $(p_2+p_3)/2$ in Eq.~(\ref{enred}) is independent of the capillary pressure drop of that burst and therefore can be averaged independently of $\Delta p$. The value of that average is $\langle p \rangle $. This assumption can be readily verified experimentally if we look at the typical temporal signature of the capillary pressure in one experiment, see Fig.~\ref{pressure_signal}. The pressure signal in this figure corresponds to an experiment performed with a liquid withdrawal rate of $1ml/h$ on a horizontal model, i.e, the angle $\theta=0^{\circ}$ (see Fig.~\ref{syst}). The data is shown after the transient regime in which the capillary pressure grows from 0 to the fluctuating signal around a characteristic pressure $\langle p \rangle$. Notice that the typical size of the fluctuations in the pressure do not seem to depend strongly on the instantaneous value of the pressure, thus confirming the assumption made in the derivation of Eq.~(\ref{work2}). This can be better seen from the insets in the shaded red and green regions in the plot. We observe the typical Haines jumps signature characteristic of slow drainage processes. The signal on the left inset is statistically very similar to that on the right, thus further justifying the assumptions in Eq.~(\ref{work2})), i.e., that the capillary pressure drop $\Delta p$ in a given burst does not depend strongly on the average pressure in the burst at steady state. The thick horizontal line in the plot denotes a characteristic pressure $\langle p \rangle =210 Pa$.

By inserting the expression for ${\cal F}(\Delta p)$ from Eq.~(\ref{fdist}) into  Eq.~(\ref{work2}), we have

\begin{multline}
  W=M \langle p \rangle  \bar{v} (\frac{n \kappa}{\bar{v}})^{2-\tau} \int_0^{\infty}  \Delta p^{1-\tau} H(\frac{\Delta p}{\Delta p^*})  d\Delta p = \\
  M  \langle p \rangle  \bar{v} \left( \frac{n \kappa}{\bar{v}}\right) ^{2-\tau} {\Delta p^*}^{2-\tau} I \; ,
\end{multline}
where $I$ is the integral 
\begin{equation}
    I=\int_0^{\infty}  y^{1-\tau} H(y)  dy \; .
    \label{int}
\end{equation}
Then by using Eq.~(\ref{pstar}) we obtain
\begin{equation}
W=M \bar{v} \langle p \rangle  I  \left(\frac{n \kappa}{\bar{v} N(p_c)} \right) ^{\frac{(2-\tau) \nu D}{1+\nu D}}  \; .
  \label{work3}
\end{equation}
Assume that we consider the time $t_w$ the front needs to move a length $w$ corresponding to the width of the model. Within this time the invading fluid has invaded an area $a^2S$ of $S$ pores.  The total number of bursts considered $M$ can be calculated as $S/\langle s \rangle $, where $ \langle s \rangle $ is the average size of the bursts  
\begin{equation}
     \langle s \rangle =I {s^*}^{2-\tau} = I \left( \frac{\kappa n}{ N(p_c) \bar{v}} \right) ^{\frac{(2-\tau)\nu D}{1+\nu D}}  \;. 
\end{equation}
where $I$ is the integral in (\ref{int}). We then get the following simple expression for the work
\begin{equation}
W= S \bar{v} \langle p \rangle  \;.
\label{work5}
\end{equation}
This expression is  the  well-known relation  of the  work expressed as  a pressure times a volume change $\langle p \rangle  \Delta V$,  where $\Delta V=S \bar{v}= S a^2 b=qt_w$ is the volume injected by the pump during the time $t_w$.

\begin{figure}
\centering
\includegraphics[width=1\linewidth]{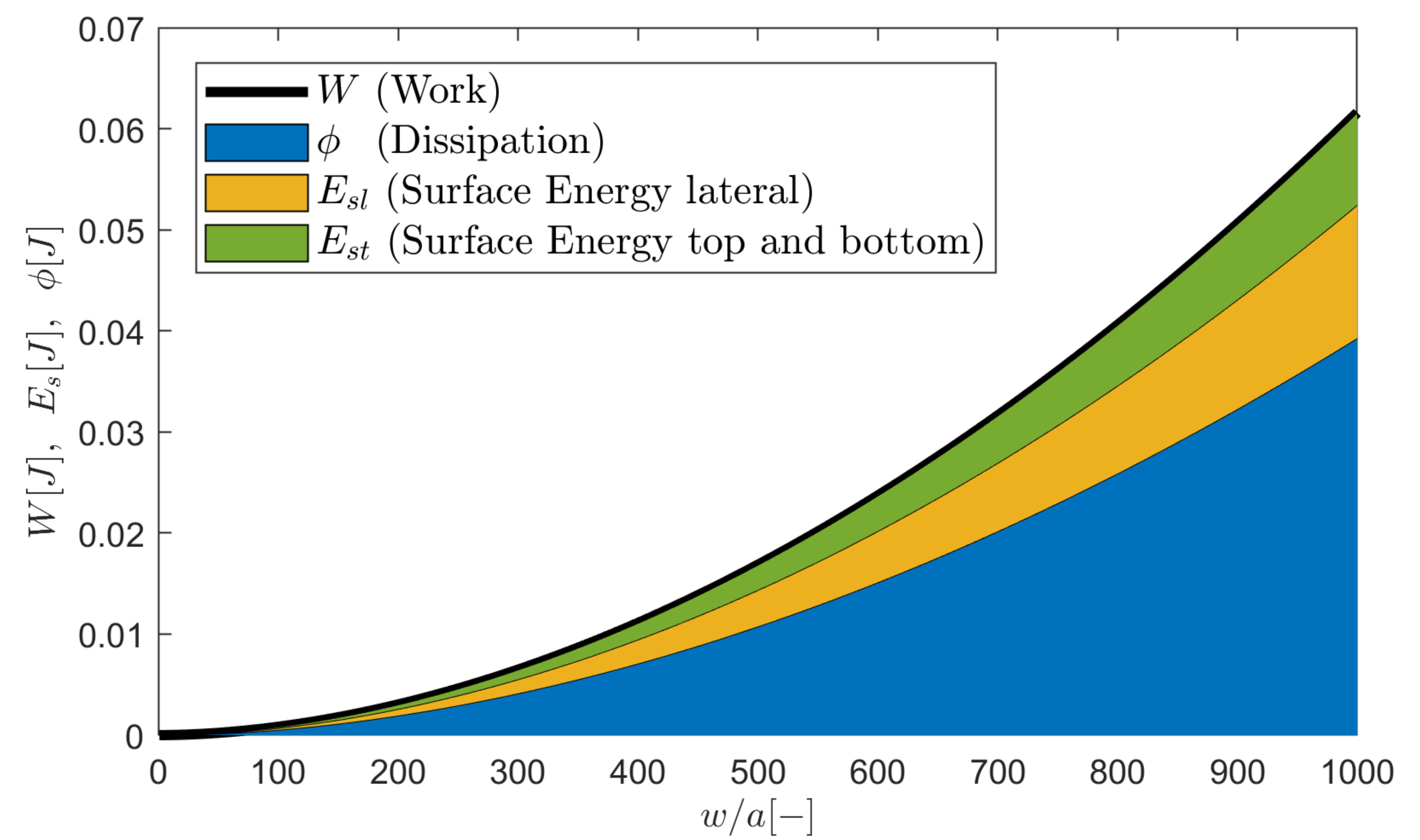}
\caption{Energy balance showing how the width of the model $w/a$ (horizontal axis) influences the applied work (thick black line), and how this work is split into energy dissipation (blue), surface energy associated to the top and bottom surfaces (green) and surface energy associated to the lateral surfaces (yellow).}
\label{energetics5} 
\end{figure}

We can then calculate the total dissipation within $t_w$  by using equations (\ref{ebudget}) and (\ref{work5})
\begin{multline}
    \Phi=W-E_s= \\
    S a^2 b \langle p \rangle-\gamma_2 a^2 2S-C \gamma_2 a b 2 S^{(D-1)/D}-\gamma_1 a b S_e \; , 
    \label{dissip13}
\end{multline}
where $C$ is a constant.  Here the second term corresponds to the contribution to the surface energy from the top and the bottom interfaces (see discussion after Eq.~(\ref{enred})) and the third term to the contribution from the side walls. The $D-1$ term in the exponent is due to the cut between the fractal invasion structure and the sidewall using one of  Mandelbrot's rules of thumb \cite{feder1988,mandelbrot1982}. The last term corresponds to the surface energy connected to the contours of the trapped clusters $S_e$  seen in red in Fig.~\ref{fig:fract_De}.  For large $S$ the  third term can be neglected and 
\begin{equation}
    \Phi \approx S a^2 b \langle p \rangle-\gamma_2 a^2 2S-\gamma_1 a b S_e \; .
    \label{approx_dissip}
\end{equation}
 The length of the front  $n$, seen as the green line in Fig.~\ref{fig:fract_De}, is  given by
\begin{equation}
    n \propto (w/a)^{D_f} \propto S^{D_f/D} \propto S_e^{D_f/D_e}\; ,
    \label{n_wa}
\end{equation}
where $D_f$ is the fractal dimension of the front (not included trapping) and $D_e$ is the fractal dimension of contours of the trapped clusters. From Fig.~\ref{fig:fract_De} we see that the measured fractal dimensions for $D_e$ and $D$ seem very close, effectively indistinguishable from one another considered the error bars. This result implies that $S \propto S_e$. In the experiment shown in Fig.~\ref{fig:fract_De} we have measured the proportionality relation to be  $S_e=1.32 S$. We have also measured the prefactor linking $S$ and the window size $w/a$ (see relation (\ref{n_wa})), obtaining the expression $S=0.72(w/a)^D$. By plugging these two equations into Eq.~(\ref{approx_dissip}) for the dissipation $\phi$, we gain access to the full energy balance of the system, the resulting plot is shown in Fig.~\ref{energetics5}. In this figure we see how the total applied work $W$ (thick black line) is split into the dissipated energy $\phi$ (blue) and the surface energy associated to creation of the top and bottom interfaces $E_{st}=\gamma_2 a^2 2S$ (green) and the lateral interface $E_{sl}=\gamma_1 a b S_e$ (yellow). The dissipation is calculated directly from Eq.~(\ref{approx_dissip}). The total applied work, the surface energy associated to the top and bottom interfaces and that associated to the lateral interfaces are given respectively by the first, second and third terms in Eq.~(\ref{approx_dissip}). Because  all terms in (\ref{approx_dissip}) and (\ref{work5}) are proportional to $S$ the ratios $\phi/W$, $E_{sl}/W$ and $E_{st}/W$ are independent of $w/a$ for large systems.

In the computation of the energy balance in Fig.~\ref{energetics5} using Eqs.~(\ref{approx_dissip}) and (\ref{n_wa}), we used $\epsilon=0.71$ measured from the experimental images (see Fig.~\ref{fig:greenorange}), $\gamma=0.064N/m$, $\psi=70^{\circ}$, $\gamma_1=(1-\epsilon)\gamma +\epsilon\gamma\cos(\psi) = 0.034N/m$, $\langle p \rangle = 210Pa$ and $D=1.87$, see Fig.~\ref{fig:fract_De}. As previously stated, our assumptions for $\gamma_1$ and $\gamma_2$ correspond to the scenario in which no film of the defending wetting fluid is left behind covering the solid surfaces after the invasion by the nonwetting phase.

If instead of the quasi-two-dimensional system considered thus far we had a three-dimensional long square channel with a cross sectional area of $w^2$, Eq.~(\ref{work5}) for the work would remain the same, while Eq.~(\ref{dissip13}) for the dissipation would be modified to 
\begin{equation}
  \Phi=S a^3 \langle p \rangle - \gamma_1 a^2 S_e -4 C \gamma_2 a^2S^{(D-1)/D}  \; ,
     \label{diss3d}
\end{equation}
where $C$ is a constant.  When the system gets large the last term can be neglected compared to the two first terms such that 
\begin{equation}
\Phi \approx S a^3 \langle p \rangle - \gamma_1 a^2 S_e \; .
     \label{Diss3d}
\end{equation}
If the invaded surface area  $S_e a^2$ times the characteristic pore size $a$ is proportional  to the invaded volume $S a^3$, the ratio $\phi/W$ and $E_{sl}/W$ will be independent of $w/a$.

\section{\label{sec:level3} Stabilizing fields and crossover lengths }
\begin{figure}
\centering
\includegraphics[width=1.0\linewidth]{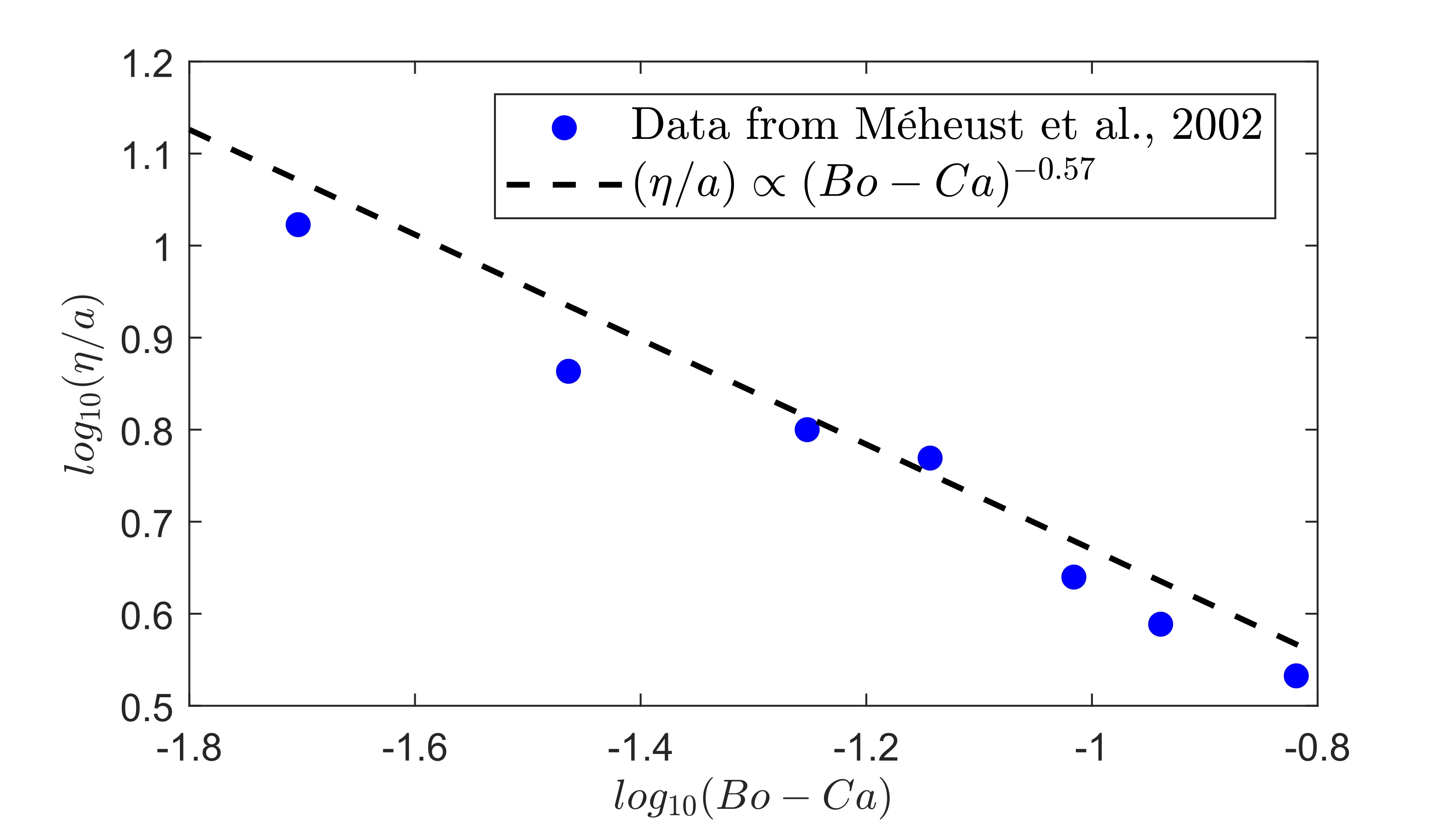}
\caption{The dependence of the front width $\eta/a$ on the generalized Bond number
$F a/(\gamma N(p_c))=Bo-Ca$. The blue dots correspond to data taken from \cite{meheust2002} where the Bond number $Bo=\Delta \rho g a^2/\gamma=0.154$ is kept fixed but the capillary number $Ca=q \mu_2 a^2/(\gamma k A)$ is changed by varying the flow rate $q$.}
\label{figYves} 
\end{figure}

Let's now take into account a gravitational field and consider an experiment  where  a low density fluid with density $\rho_1$ and viscosity $\mu_1=0$ invades another  fluid with a higher density $\rho_2$  and viscosity $\mu_2$ from above. (See Fig.~\ref{syst}). 
The mapping between the occupation probability in percolation theory $f$ and the capillary pressure $p$ is given by \cite{roux1989,stauffer1994}
\begin{equation}
  f=\int^{p}_0 N(p_t) dp_t  \; . 
  \label{percmap}
\end{equation}
We therefore have
\begin{equation}
  f-f_c=\int^{p}_{p_c} N(p_t) dp_t  \; ,
  \label{intt3}
\end{equation}
where the  critical occupation probability is $f_c$ and the critical percolation pressure is $p_c$. As before, $N(p_t)$ is the normalized capillary pressure threshold  distribution. Now, by Taylor expanding $N(p_t)$ around $p_c$ to the lowest order in $p-p_c$ in Eq.~(\ref{intt3}) we get
\begin{equation}
  f-f_c=N(p_c)(p-p_c) \; .
  \label{int3}
\end{equation}
It is reasonable to assume that the viscous pressure drop in the displaced fluid will depend  linearly on the length scale since there are no trapped invading fluid clusters in the displaced fluid.  Let us consider the capillary pressure at a height $x$.  Since the gravitational field  also depends linearly on the length scale $x$ we have
\begin{equation}
  f-f_c=N(p_c)(\Delta \rho g a - \frac{q \mu_2 a}{kA})(x-x_0)/a=F (x-x_0)/a\; ,
  \label{fint3}
\end{equation}
where $x_0$ corresponds to a the height   along the front where the capillary pressure is at the percolation threshold. 
Here $k$ is the permeability felt by the displaced fluid, $A$ the cross-section area of the porous medium, and $\Delta \rho= \rho_2-\rho_1$ the density difference between the fluids.  The fluctuation number $F$
\begin{equation} 
    F=N(p_c) (\Delta \rho g a - \frac{q \mu_2 a}{kA}) \; , 
    \label{fluc0}
\end{equation}
is a dimensionless number \cite{meheust2002} which characterizes the gravitational and viscous fields and the capillary pressure fluctuations.  Note that the  width of the capillary pressure threshold distribution is  $\sigma \approx 1/N(p_c)$, as the distribution is normalized. 
We will then use an assumption, first introduced by Sapoval \cite{sapova1985,gouyet1988,wilkinson1983,birovljev91}, that the correlation length in percolation  will scale  as the width of the front $\eta=(x-x_0) \propto \xi$, where $\xi$ is the percolation correlation length. This is based on the observation that the largest trapped cluster of wetting fluid is limited by the front width. From percolation theory, the correlation length $\xi$ is given by \cite{stauffer1994}
\begin{equation}
    (\xi/a) \propto (f-f_c)^{-\nu} \; ,
    \label{corrl}
\end{equation}
Using Sapoval's argument, and inserting Eq.~(\ref{corrl}) into Eq.~(\ref{fint3}) gives \cite{wilkinson1984,meheust2002}
 \begin{equation}
  \eta/a \propto F^{\frac{-\nu}{1+\nu}} \; .
  \label{fluc}
\end{equation}
Therefore, when $F>0$, the front will be stabilized with a characteristic  length scale $\eta$  \cite{wilkinson1984,birovljev91,meheust2002} . This scaling behaviour is shown in Fig.~\ref{figYves}. The blue dots show experimental data taken from \cite{meheust2002} where the authors analyzed the scaling of the invasion front in drainage by keeping the Bond number $Bo=\Delta \rho g a^2/\gamma=0.154$ fixed but changing the capillary number $Ca$ by varying the flow rate $q$ (see Fig. 10 in Ref.~\cite{meheust2002} and experiments 1, 2, 3, 4, 5, 7, and 9 in Table I in the same reference).

Let us first assume that the gravitational effect is large enough such that it  is the characteristic length scale $\eta$ that sets the width of the front $\eta<w$. Consider again a time scale $t_w$ corresponding to the  time the front needs to move a length scale $w$.
Dividing space into cells of size $\eta^2$, below which the fluid distribution is fractal, 
the number of pores $S$ and $S_e$ are given by 
\begin{equation}
    S \propto (\frac{w}{\eta})^2 (\frac{\eta}{a})^D \; , \; 
    S_e \propto (\frac{w}{\eta})^2 (\frac{\eta}{a})^{D_e} \; , 
\end{equation}
and the  work can be written as 
\begin{multline}
   W=-g(\rho_2 h +\rho_1 (L-h))qt_w+ \frac{q \mu_2 h}{k A} S a^2 b\\+S a^2 b \langle p \rangle \; ,
   \label{work4}
\end{multline}
where $h$ is the average position of the front from the lower outlet. The first term in (\ref{work4}) is the negative work due to the hydrostatic pressure, the second term the work due to the  viscous pressure in the displaced fluid,  and the last term is the work needed to continuously build up the interface energy. 
The work must be equal to the dissipation $\Phi$ plus the surface energy  $E_s$ minus the change in gravitational potential energy $\Delta U_p$
(corresponding to the first term in Eq.~(\ref{work4}))
\begin{equation}
    W=\Phi+E_s-\Delta U_p \; .
\end{equation}
We then get the dissipation
\begin{multline}
\Phi=\frac{q \mu_2 h}{\kappa A} S a^2 b + S a^2 b \langle p \rangle  \\ - \gamma_2 a^2 2 S-C \gamma_2 a b  2 S^{(D-1)/D}-  \gamma_1 a b S_e^{D_e} \;  ,
\label{dissip_visc_1}
\end{multline}
which for large S can be approximated as 
\begin{equation}
\Phi=\frac{q \mu_2 h}{\kappa A} S a^2 b+ S a^2 b \langle p \rangle  - \gamma_2 a^2 2 S-  \gamma_1 a b S_e^{D_e} \;  .
\label{dissip_visc_2}
\end{equation}
Note that in the limit of very slow processes, the flow rate $q$ becomes negligible so the first term in this Eqs.~(\ref{dissip_visc_1}) and (\ref{dissip_visc_2}) can be ignored, thus recovering Eqs.~(\ref{dissip13}) and (\ref{approx_dissip}). In the scenario in which $\eta>w$, the relations for $S$  and $S_e$ will change to
\begin{equation}
    S \propto (\frac{w}{a})^D \; , \; 
    S_e \propto (\frac{w}{a})^{D_e} \;  ,
\end{equation}
but we will still have the same expressions for $W$ and $\phi$.  
Again if $D_e=D$, the ratios $\phi/W$,   $E_{sl}/W$ and $E_{st}/W$  will be independent of $w/a$ for large systems.

\section{\label{sec:level4} Saturation behind the invasion front}
\begin{figure}
\centering
\includegraphics[width=1.0\linewidth]{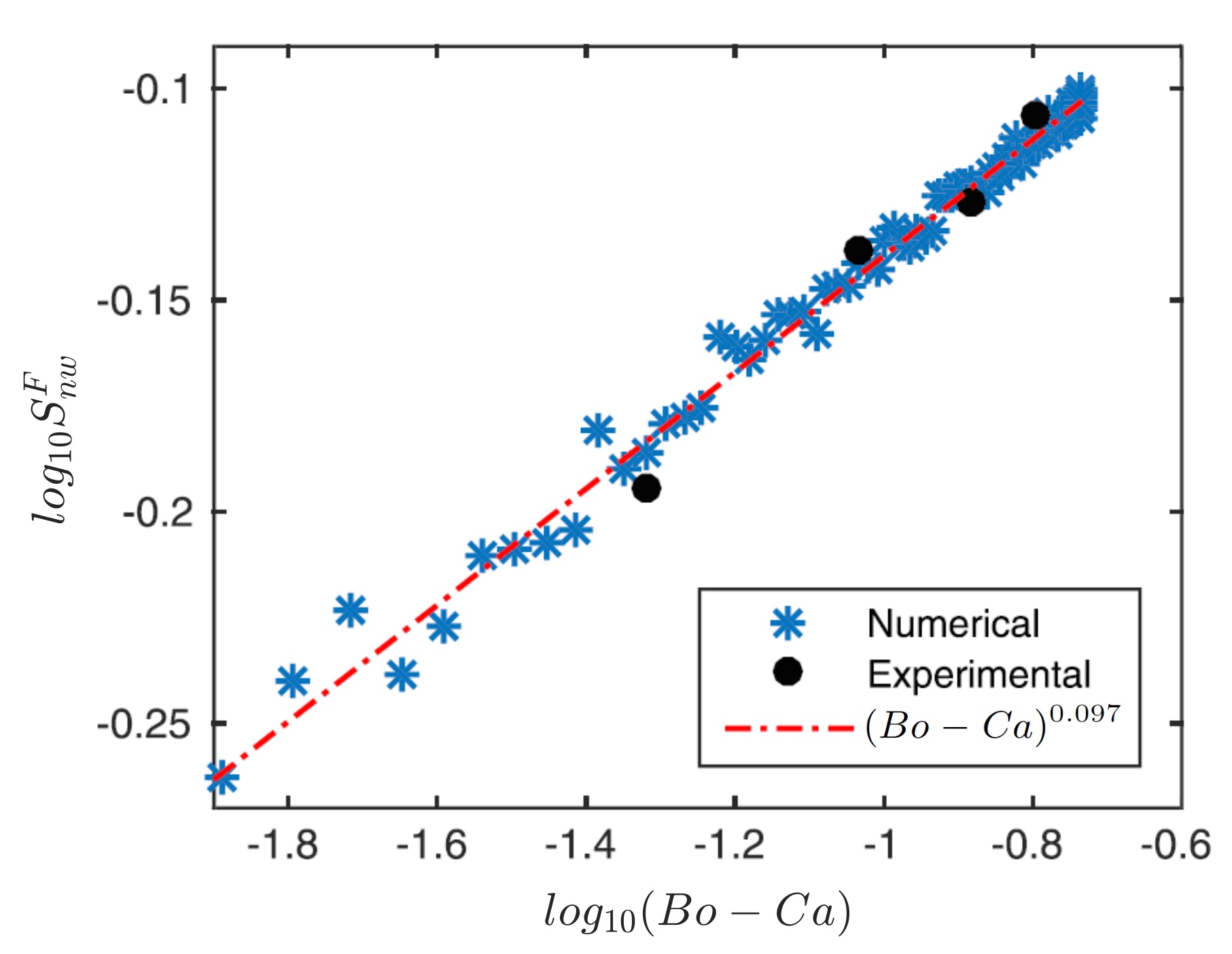}
\caption{The final saturation as  a function of the generalized Bond number $F a/(\gamma N(p_c))=Bo-Ca$ is plotted both for experimental (black dots) and numerical results (blue stars) produced via invasion percolation with a gravitational field, together with the predicted result Eq.~(\ref{sateq}) (red dash-dotted line) in two dimensions. The capillary number $Ca=q \mu_2 a^2/(\gamma k A)=1.2 \cdot 10^{-4}$ is kept fixed while the Bond number $Bo=\Delta \rho g a^2/\gamma$ is changed in the experiments.  Numerical and experimental data from \cite{ayaz2020}.}
\label{satfig} 
\end{figure}

The upscaling problem of calculating the large scale saturation involves identifying the cross-over length scale where the fluid distribution is no longer fractal.  Let $L$  be the length of the porous model and assume $L>w$.  The final  saturation behind the front of the invading fluid  and its dependence  on the pressure across the model has been studied in Ref.~\cite{ayaz2020}.  In this study, we   considered   the volume of the invaded fluid  in boxes with size corresponding to the width of the invading front. On length scales below this size, the structure of the invading fluid is fractal, while on  length scales larger than this size, the structure is homogeneous.  For sufficiently large $F$, when $\eta$ is the characteristic length scale of the front, the saturation $S^F_{nw}$ of the nonwetting fluid behind the front becomes \cite{ayaz2020}
\begin{equation}
S^F_{nw} \propto \frac{(\frac{Lw^{d-1}}{\eta^d})(\frac{\eta}{a})^D a^ d}{Lw^{d-1}}=(\frac{\eta}{a})^{D-d} \; ,
\label{sat1}
\end{equation}
where $d$ is the spatial dimension (2 or 3). Hence using Eq.~(\ref{fluc})
\begin{equation}
  S^F_{nw} \propto  F^{-\nu (D-d)/(\nu+1)}  \; .
\label{sateq}
\end{equation}
However, when the fluctuation number  $F$ is sufficient small, $\eta>w$,  and the width of the model $w$ will be the characteristic length scale in the problem. Then 
\begin{equation}
S^F_{nw} \propto (\frac{w}{a})^{(D-d)}   \; .
\end{equation}
Fig.~\ref{satfig} shows a two-dimensional invasion percolation simulation with a gravitational field together with drainage experiments performed by Ayaz et al. \cite{ayaz2020}. The red dash-dotted line confirms the predictions of the theoretical scaling in Eq.~(\ref{sateq}). We make the important remark here that in the equations considered above  we have neglected the boundary effects at the inlet and outlet since $L/w \gg 1$. These boundary effects are however important when the characteristic length scale of the front $w$ is of the same order as the length $L$ of the system \cite{moura2015}.

In section IV we considered an invading fluid with  a negligible  viscosity $\mu_1=0$.  Let us now instead consider the scenario in which this fluid has  a non-negligible  viscosity $\mu_1$. In that case the fluctuation number will be 
\begin{equation} 
    F=N(p_c) (\Delta \rho g a + \frac{q \mu_1 a}{k_1 A} - \frac{q \mu_2 a}{k A}) \; , 
    \label{fluc1}
\end{equation}
where $k_1$ is an effective permeability that depends on the characteristic crossover length scale of the system which is either $\eta$ or $w$. 
The expression (\ref{fluc1}) is meaningful only if the pressure drop in the invading fluid depends linearly on the length scale along the average flow direction. Network simulations  in two dimensions however,   show   a linear change  in the capillary pressure  due to viscous flow \cite{aker2000B}, \cite{aker2000C}. This   is due to the loop-less strands  in the invading nonwetting fluid.  It is of great interest to know if this is also valid  in  three dimensions.  

The structure left behind the front is a  fractal structure with clusters of all length scales up to the characteristic length scale $\eta $ or $w$ \cite{birovljev91,frette97}. Therefore if $\eta<w$, the permeability $k_1(F)$ will be a function of $F$. If we know this functional dependence  we can solve  the equation 
\begin{equation} 
    F=N(p_c) (\Delta \rho g a +\frac{q \mu_1 a}{k_1(F) A} - \frac{q \mu_2 a}{k A}) \; , 
    \label{fluc2}
\end{equation}
with respect to $F$ and thereby find the characteristic length scale $\eta$ from  
\begin{equation}
  \eta/a \propto F^{\frac{-\nu}{1+\nu}}  \; . 
\end{equation}
We then also know the saturation through  Eq.~(\ref{sat1}).
A goal for future experiments/simulations/theory should  therefore be to measure/predict the permeability dependence of the invading  fluid as function of the characteristic length scale $\eta$.  \\ \\

\section{\label{sec:level5} Conclusion}
In this paper, we discussed the importance of capillary fluctuations in porous media, as well as the characteristic length scales set by the competition between capillary fluctuations and external fields, gravitational or viscous.  We focused on fluid fronts that are stabilized by gravitational and viscous fields. The fluctuation number $F$, which describes the scaling of the front width $\eta$, was introduced.  When considering viscous and gravitational fields, the theory describes well the scaling of the width of the fluid front and the final saturation of the fluid left behind the invasion front observed in experiments. On a length scale smaller than $\eta$, the structure within the front is generally fractal, while on a length scale larger than $\eta$, it is homogeneous. The characteristic length scale $\eta$ is thus of primary importance in defining a relevant Representative Elementary Volume (REV) for an average Darcy description of the two-phase flow problem.  

 We also discussed the energy dissipation caused by out-of-equilibrium Haines jumps and calculated the total elastic energy released by those jumps using the pressure drops connected to the jumps' known scaling behavior.  When the invasion front is in a steady state regime, we find that the elastic energy released by the jumps corresponds to the work applied to the system, as expected.
However, this description also explains the local activity and dissipation caused by local Haines jumps, as well as how it is related to dissipation on a larger scale. 

We calculated the energy dissipation and saturation left behind the invasion front in addition to the work done by the pump to push the fluid front forward. These quantities are affected by the generalized fluctuation number, as well as the fractal dimension of the invasion structure, the fractal dimension of the contour of the trapped clusters, and the exponent $\nu$, which describes the divergence of the correlation length in percolation as it approaches the critical percolation threshold $f_c$.  The work $W$, dissipation $\Phi$, and surface energy $E_s$ were discovered to be proportional to the number of pores invaded $S$, with fractal scaling on length scales smaller than the characteristic length scale, $\eta$ or $w$, but scaling with the spatial dimension on larger length scales. Hence, the ratio of these energies stays constant as the system size increases.

The theory described here is a bottom-up approach that allows the problem to be scaled up from small to large scales. The theoretical results were compared to and found to be consistent with quasi-two-dimensional experiments. Because of the small temperature increase involved in heat dissipation from the Haines jumps (mK or less), it is very challenging to measure these temperature fluctuations. Such a measurement remains to be done. An experimental verification of the theoretical scaling of the invasion front and the saturation behind the front for three-dimensional porous media is also of great interest for future work \cite{ayaz2020}.  Experiments involving changing the gravitational effects in a geophysical centrifuge could be one way to accomplish this. The porous medium in a real rock is typically not homogeneous, in contrast to the porous media considered here.  As a result, it is of great interest to extend the theory to include also the case of inhomogeneous porous media, for example with a porosity gradient.  \\ \\ \\
{\bf ACKNOWLEDGMENTS } \\
\noindent This work was partly supported by the Research Council of Norway through its Centers of Excellence funding scheme, project number 262644. 
We thank Einar Hinrichsen, Jens Feder, Torstein J{\o}ssang, Amnon Aharony, Paul Meakin, Jean Schmittbuhl, Gerhard Sch{\"a}fer, Liv Furuberg, Alexander Birovljev, Yves Meheust, Per Arne Rikvold, Harold Auradou,  Olav Inge Frette, Vidar Frette, Grunde L{\o}voll, Bj{\o}rnar Sandnes, Mihailo Jankov, Fredrik Kvalheim Eriksen, Monem Ayaz, and Tom Vincent-Dospital  for fruitful discussions.


\begin{thebibliography}{81}%
\makeatletter
\providecommand \@ifxundefined [1]{%
 \@ifx{#1\undefined}
}%
\providecommand \@ifnum [1]{%
 \ifnum #1\expandafter \@firstoftwo
 \else \expandafter \@secondoftwo
 \fi
}%
\providecommand \@ifx [1]{%
 \ifx #1\expandafter \@firstoftwo
 \else \expandafter \@secondoftwo
 \fi
}%
\providecommand \natexlab [1]{#1}%
\providecommand \enquote  [1]{``#1''}%
\providecommand \bibnamefont  [1]{#1}%
\providecommand \bibfnamefont [1]{#1}%
\providecommand \citenamefont [1]{#1}%
\providecommand \href@noop [0]{\@secondoftwo}%
\providecommand \href [0]{\begingroup \@sanitize@url \@href}%
\providecommand \@href[1]{\@@startlink{#1}\@@href}%
\providecommand \@@href[1]{\endgroup#1\@@endlink}%
\providecommand \@sanitize@url [0]{\catcode `\\12\catcode `\$12\catcode
  `\&12\catcode `\#12\catcode `\^12\catcode `\_12\catcode `\%12\relax}%
\providecommand \@@startlink[1]{}%
\providecommand \@@endlink[0]{}%
\providecommand \url  [0]{\begingroup\@sanitize@url \@url }%
\providecommand \@url [1]{\endgroup\@href {#1}{\urlprefix }}%
\providecommand \urlprefix  [0]{URL }%
\providecommand \Eprint [0]{\href }%
\providecommand \doibase [0]{https://doi.org/}%
\providecommand \selectlanguage [0]{\@gobble}%
\providecommand \bibinfo  [0]{\@secondoftwo}%
\providecommand \bibfield  [0]{\@secondoftwo}%
\providecommand \translation [1]{[#1]}%
\providecommand \BibitemOpen [0]{}%
\providecommand \bibitemStop [0]{}%
\providecommand \bibitemNoStop [0]{.\EOS\space}%
\providecommand \EOS [0]{\spacefactor3000\relax}%
\providecommand \BibitemShut  [1]{\csname bibitem#1\endcsname}%
\let\auto@bib@innerbib\@empty
%</preamble>
\bibitem [{\citenamefont {Anderson}\ \emph {et~al.}(2010)\citenamefont
  {Anderson}, \citenamefont {Zhang}, \citenamefont {Ding}, \citenamefont
  {Blanco}, \citenamefont {Bi},\ and\ \citenamefont
  {Wilkinson}}]{anderson2010}%
  \BibitemOpen
  \bibfield  {author} {\bibinfo {author} {\bibfnamefont {R.}~\bibnamefont
  {Anderson}}, \bibinfo {author} {\bibfnamefont {L.}~\bibnamefont {Zhang}},
  \bibinfo {author} {\bibfnamefont {Y.}~\bibnamefont {Ding}}, \bibinfo {author}
  {\bibfnamefont {M.}~\bibnamefont {Blanco}}, \bibinfo {author} {\bibfnamefont
  {X.}~\bibnamefont {Bi}},\ and\ \bibinfo {author} {\bibfnamefont {D.~P.}\
  \bibnamefont {Wilkinson}},\ }\bibfield  {title} {\bibinfo {title} {A critical
  review of two-phase flow in gas flow channels of proton exchange membrane
  fuel cells},\ }\href
  {https://doi.org/https://doi.org/10.1016/j.jpowsour.2009.12.123} {\bibfield
  {journal} {\bibinfo  {journal} {Journal of Power Sources}\ }\textbf {\bibinfo
  {volume} {195}},\ \bibinfo {pages} {4531 } (\bibinfo {year}
  {2010})}\BibitemShut {NoStop}%
\bibitem [{\citenamefont {Hwang}\ \emph {et~al.}(2012)\citenamefont {Hwang},
  \citenamefont {Kim}, \citenamefont {Song}, \citenamefont {Son},\ and\
  \citenamefont {Jang}}]{hwang2012}%
  \BibitemOpen
  \bibfield  {author} {\bibinfo {author} {\bibfnamefont {S.~H.}\ \bibnamefont
  {Hwang}}, \bibinfo {author} {\bibfnamefont {C.}~\bibnamefont {Kim}}, \bibinfo
  {author} {\bibfnamefont {H.}~\bibnamefont {Song}}, \bibinfo {author}
  {\bibfnamefont {S.}~\bibnamefont {Son}},\ and\ \bibinfo {author}
  {\bibfnamefont {J.}~\bibnamefont {Jang}},\ }\bibfield  {title} {\bibinfo
  {title} {Designed architecture of multiscale porous tio2 nanofibers for
  dye-sensitized solar cells photoanode},\ }\href
  {https://doi.org/10.1021/am301245s} {\bibfield  {journal} {\bibinfo
  {journal} {ACS Appl. Mater. Interfaces}\ }\textbf {\bibinfo {volume} {4}},\
  \bibinfo {pages} {5287} (\bibinfo {year} {2012})}\BibitemShut {NoStop}%
\bibitem [{\citenamefont {Cantwell}\ and\ \citenamefont
  {Morton}(1991)}]{cantwell1991}%
  \BibitemOpen
  \bibfield  {author} {\bibinfo {author} {\bibfnamefont {W.}~\bibnamefont
  {Cantwell}}\ and\ \bibinfo {author} {\bibfnamefont {J.}~\bibnamefont
  {Morton}},\ }\bibfield  {title} {\bibinfo {title} {The impact resistance of
  composite materials — a review},\ }\href
  {https://doi.org/https://doi.org/10.1016/0010-4361(91)90549-V} {\bibfield
  {journal} {\bibinfo  {journal} {Composites}\ }\textbf {\bibinfo {volume}
  {22}},\ \bibinfo {pages} {347} (\bibinfo {year} {1991})}\BibitemShut
  {NoStop}%
\bibitem [{\citenamefont {Gibson}\ \emph {et~al.}(1999)\citenamefont {Gibson},
  \citenamefont {Rivin}, \citenamefont {Kendrick},\ and\ \citenamefont
  {Schreuder-Gibson}}]{gibson1999}%
  \BibitemOpen
  \bibfield  {author} {\bibinfo {author} {\bibfnamefont {P.}~\bibnamefont
  {Gibson}}, \bibinfo {author} {\bibfnamefont {D.}~\bibnamefont {Rivin}},
  \bibinfo {author} {\bibfnamefont {C.}~\bibnamefont {Kendrick}},\ and\
  \bibinfo {author} {\bibfnamefont {H.}~\bibnamefont {Schreuder-Gibson}},\
  }\bibfield  {title} {\bibinfo {title} {Humidity-dependent air permeability of
  textile materials1},\ }\href {https://doi.org/10.1177/004051759906900501}
  {\bibfield  {journal} {\bibinfo  {journal} {Textile Research Journal}\
  }\textbf {\bibinfo {volume} {69}},\ \bibinfo {pages} {311} (\bibinfo {year}
  {1999})},\ \Eprint
  {https://arxiv.org/abs/https://doi.org/10.1177/004051759906900501}
  {https://doi.org/10.1177/004051759906900501} \BibitemShut {NoStop}%
\bibitem [{\citenamefont {Lake}(1989)}]{lake1989}%
  \BibitemOpen
  \bibfield  {author} {\bibinfo {author} {\bibfnamefont {L.}~\bibnamefont
  {Lake}},\ }\href@noop {} {\emph {\bibinfo {title} {Enhanced oil recovery}}}\
  (\bibinfo  {publisher} {Prentice Hall},\ \bibinfo {address} {Englewood
  Cliffs, N.J},\ \bibinfo {year} {1989})\BibitemShut {NoStop}%
\bibitem [{\citenamefont {Shabani~Afrapoli}\ \emph {et~al.}(2011)\citenamefont
  {Shabani~Afrapoli}, \citenamefont {Alipour},\ and\ \citenamefont
  {Tors\ae{}ter}}]{afrapoli2011}%
  \BibitemOpen
  \bibfield  {author} {\bibinfo {author} {\bibfnamefont {M.}~\bibnamefont
  {Shabani~Afrapoli}}, \bibinfo {author} {\bibfnamefont {S.}~\bibnamefont
  {Alipour}},\ and\ \bibinfo {author} {\bibfnamefont {O.}~\bibnamefont
  {Tors\ae{}ter}},\ }\bibfield  {title} {\bibinfo {title} {Fundamental study of
  pore scale mechanisms in microbial improved oil recovery processes},\ }\href
  {https://doi.org/10.1007/s11242-011-9825-7} {\bibfield  {journal} {\bibinfo
  {journal} {Transport in Porous Media}\ }\textbf {\bibinfo {volume} {90}},\
  \bibinfo {pages} {949} (\bibinfo {year} {2011})}\BibitemShut {NoStop}%
\bibitem [{\citenamefont {Yan}\ \emph {et~al.}(2012)\citenamefont {Yan},
  \citenamefont {Luo}, \citenamefont {Wang}, \citenamefont {Toussaint},
  \citenamefont {Schmittbuhl}, \citenamefont {Vasseur}, \citenamefont {Chen},
  \citenamefont {Yu},\ and\ \citenamefont {Zhang}}]{yan2012}%
  \BibitemOpen
  \bibfield  {author} {\bibinfo {author} {\bibfnamefont {J.}~\bibnamefont
  {Yan}}, \bibinfo {author} {\bibfnamefont {X.}~\bibnamefont {Luo}}, \bibinfo
  {author} {\bibfnamefont {W.}~\bibnamefont {Wang}}, \bibinfo {author}
  {\bibfnamefont {R.}~\bibnamefont {Toussaint}}, \bibinfo {author}
  {\bibfnamefont {J.}~\bibnamefont {Schmittbuhl}}, \bibinfo {author}
  {\bibfnamefont {G.}~\bibnamefont {Vasseur}}, \bibinfo {author} {\bibfnamefont
  {F.}~\bibnamefont {Chen}}, \bibinfo {author} {\bibfnamefont {A.}~\bibnamefont
  {Yu}},\ and\ \bibinfo {author} {\bibfnamefont {L.}~\bibnamefont {Zhang}},\
  }\bibfield  {title} {\bibinfo {title} {An experimental study of secondary oil
  migration in a three-dimensional tilted porous medium},\ }\href
  {https://doi.org/10.1306/09091110140} {\bibfield  {journal} {\bibinfo
  {journal} {{AAPG} bulletin}\ }\textbf {\bibinfo {volume} {96}},\ \bibinfo
  {pages} {773} (\bibinfo {year} {2012})}\BibitemShut {NoStop}%
\bibitem [{\citenamefont {Vasseur}\ \emph {et~al.}(2013)\citenamefont
  {Vasseur}, \citenamefont {Luo}, \citenamefont {Yan}, \citenamefont {Loggia},
  \citenamefont {Toussaint},\ and\ \citenamefont {Schmittbuhl}}]{vasseur2013}%
  \BibitemOpen
  \bibfield  {author} {\bibinfo {author} {\bibfnamefont {G.}~\bibnamefont
  {Vasseur}}, \bibinfo {author} {\bibfnamefont {X.}~\bibnamefont {Luo}},
  \bibinfo {author} {\bibfnamefont {J.}~\bibnamefont {Yan}}, \bibinfo {author}
  {\bibfnamefont {D.}~\bibnamefont {Loggia}}, \bibinfo {author} {\bibfnamefont
  {R.}~\bibnamefont {Toussaint}},\ and\ \bibinfo {author} {\bibfnamefont
  {J.}~\bibnamefont {Schmittbuhl}},\ }\bibfield  {title} {\bibinfo {title}
  {Flow regime associated with vertical secondary migration},\ }\href
  {https://doi.org/10.1016/j.marpetgeo.2013.04.020} {\bibfield  {journal}
  {\bibinfo  {journal} {Marine and Petroleum Geology}\ }\textbf {\bibinfo
  {volume} {45}},\ \bibinfo {pages} {150} (\bibinfo {year} {2013})}\BibitemShut
  {NoStop}%
\bibitem [{\citenamefont {Guymon}(1994)}]{guymon1994}%
  \BibitemOpen
  \bibfield  {author} {\bibinfo {author} {\bibfnamefont {G.}~\bibnamefont
  {Guymon}},\ }\href@noop {} {\emph {\bibinfo {title} {Unsaturated zone
  hydrology}}}\ (\bibinfo  {publisher} {Prentice Hall},\ \bibinfo {address}
  {Englewood Cliffs, N.J},\ \bibinfo {year} {1994})\BibitemShut {NoStop}%
\bibitem [{\citenamefont {Bear}(1972)}]{bear1972}%
  \BibitemOpen
  \bibfield  {author} {\bibinfo {author} {\bibfnamefont {J.}~\bibnamefont
  {Bear}},\ }\href@noop {} {\emph {\bibinfo {title} {Dynamics of fluids in
  porous media}}}\ (\bibinfo  {publisher} {Elsevier},\ \bibinfo {address} {New
  York},\ \bibinfo {year} {1972})\BibitemShut {NoStop}%
\bibitem [{\citenamefont {Bear}\ and\ \citenamefont
  {Verruijt}(1987)}]{bear1987}%
  \BibitemOpen
  \bibfield  {author} {\bibinfo {author} {\bibfnamefont {J.}~\bibnamefont
  {Bear}}\ and\ \bibinfo {author} {\bibfnamefont {A.}~\bibnamefont
  {Verruijt}},\ }\href@noop {} {\emph {\bibinfo {title} {Modeling groundwater
  flow and pollution}}}\ (\bibinfo  {publisher} {D. Reidel Pub. Co. Sold and
  distributed in the U.S.A. and Canada by Kluwer Academic Publishers},\
  \bibinfo {address} {Dordrecht Boston Norwell, MA, U.S.A},\ \bibinfo {year}
  {1987})\BibitemShut {NoStop}%
\bibitem [{\citenamefont {Jellali}\ \emph {et~al.}(2001)\citenamefont
  {Jellali}, \citenamefont {Muntzer}, \citenamefont {Razakarisoa},\ and\
  \citenamefont {Sch{\"a}fer}}]{jellali2001}%
  \BibitemOpen
  \bibfield  {author} {\bibinfo {author} {\bibfnamefont {S.}~\bibnamefont
  {Jellali}}, \bibinfo {author} {\bibfnamefont {P.}~\bibnamefont {Muntzer}},
  \bibinfo {author} {\bibfnamefont {O.}~\bibnamefont {Razakarisoa}},\ and\
  \bibinfo {author} {\bibfnamefont {G.}~\bibnamefont {Sch{\"a}fer}},\
  }\bibfield  {title} {\bibinfo {title} {Large scale experiment on transport of
  trichloroethylene in a controlled aquifer},\ }\href
  {https://doi.org/10.1023/A:1010652230922} {\bibfield  {journal} {\bibinfo
  {journal} {Transp. Porous Media}\ }\textbf {\bibinfo {volume} {44}},\
  \bibinfo {pages} {145} (\bibinfo {year} {2001})}\BibitemShut {NoStop}%
\bibitem [{\citenamefont {Nsir}\ \emph {et~al.}(2012)\citenamefont {Nsir},
  \citenamefont {Schäfer}, \citenamefont {di~Chiara~Roupert}, \citenamefont
  {Razakarisoa},\ and\ \citenamefont {Toussaint}}]{nsir2012}%
  \BibitemOpen
  \bibfield  {author} {\bibinfo {author} {\bibfnamefont {K.}~\bibnamefont
  {Nsir}}, \bibinfo {author} {\bibfnamefont {G.}~\bibnamefont {Schäfer}},
  \bibinfo {author} {\bibfnamefont {R.}~\bibnamefont {di~Chiara~Roupert}},
  \bibinfo {author} {\bibfnamefont {O.}~\bibnamefont {Razakarisoa}},\ and\
  \bibinfo {author} {\bibfnamefont {R.}~\bibnamefont {Toussaint}},\ }\bibfield
  {title} {\bibinfo {title} {Laboratory experiments on {DNAPL} gravity
  fingering in water-saturated porous media},\ }\href@noop {} {\bibfield
  {journal} {\bibinfo  {journal} {International Journal of Multiphase Flow}\
  }\textbf {\bibinfo {volume} {40}},\ \bibinfo {pages} {83 } (\bibinfo {year}
  {2012})}\BibitemShut {NoStop}%
\bibitem [{\citenamefont {Gagn{\'e}}(2021)}]{gagne2021}%
  \BibitemOpen
  \bibfield  {author} {\bibinfo {author} {\bibfnamefont {J.}~\bibnamefont
  {Gagn{\'e}}},\ }\href {https://books.google.no/books?id=Lcg8zgEACAAJ} {\emph
  {\bibinfo {title} {The Physics of Filter Coffee}}}\ (\bibinfo  {publisher}
  {Scott Rao Coffee Books},\ \bibinfo {year} {2021})\BibitemShut {NoStop}%
\bibitem [{\citenamefont {Chen}\ and\ \citenamefont
  {Wilkinson}(1985)}]{chen1985}%
  \BibitemOpen
  \bibfield  {author} {\bibinfo {author} {\bibfnamefont {J.-D.}\ \bibnamefont
  {Chen}}\ and\ \bibinfo {author} {\bibfnamefont {D.}~\bibnamefont
  {Wilkinson}},\ }\bibfield  {title} {\bibinfo {title} {Pore-scale viscous
  fingering in porous media},\ }\href
  {https://doi.org/10.1103/PhysRevLett.55.1892} {\bibfield  {journal} {\bibinfo
   {journal} {Phys. Rev. Lett.}\ }\textbf {\bibinfo {volume} {55}},\ \bibinfo
  {pages} {1892} (\bibinfo {year} {1985})}\BibitemShut {NoStop}%
\bibitem [{\citenamefont {M\aa{}l\o{}y}\ \emph {et~al.}(1985)\citenamefont
  {M\aa{}l\o{}y}, \citenamefont {Feder},\ and\ \citenamefont
  {J\o{}ssang}}]{maloy1985}%
  \BibitemOpen
  \bibfield  {author} {\bibinfo {author} {\bibfnamefont {K.~J.}\ \bibnamefont
  {M\aa{}l\o{}y}}, \bibinfo {author} {\bibfnamefont {J.}~\bibnamefont
  {Feder}},\ and\ \bibinfo {author} {\bibfnamefont {T.}~\bibnamefont
  {J\o{}ssang}},\ }\bibfield  {title} {\bibinfo {title} {Viscous fingering
  fractals in porous media},\ }\href@noop {} {\bibfield  {journal} {\bibinfo
  {journal} {Phys. Rev. Lett.}\ }\textbf {\bibinfo {volume} {55}},\ \bibinfo
  {pages} {2688–} (\bibinfo {year} {1985})}\BibitemShut {NoStop}%
\bibitem [{\citenamefont {Weitz}\ \emph {et~al.}(1987)\citenamefont {Weitz},
  \citenamefont {Stokes}, \citenamefont {Ball},\ and\ \citenamefont
  {Kushnick}}]{weitz1987}%
  \BibitemOpen
  \bibfield  {author} {\bibinfo {author} {\bibfnamefont {D.}~\bibnamefont
  {Weitz}}, \bibinfo {author} {\bibfnamefont {J.}~\bibnamefont {Stokes}},
  \bibinfo {author} {\bibfnamefont {R.}~\bibnamefont {Ball}},\ and\ \bibinfo
  {author} {\bibfnamefont {A.}~\bibnamefont {Kushnick}},\ }\bibfield  {title}
  {\bibinfo {title} {Dynamic capillary pressure in porous media: Origin of the
  viscous-fingering length scale,},\ }\href
  {https://doi.org/10.1103/PhysRevLett.59.2967} {\bibfield  {journal} {\bibinfo
   {journal} {Phys. Rev. Lett}\ }\textbf {\bibinfo {volume} {59}},\ \bibinfo
  {pages} {2967} (\bibinfo {year} {1987})}\BibitemShut {NoStop}%
\bibitem [{\citenamefont {Lenormand}\ \emph {et~al.}(1988)\citenamefont
  {Lenormand}, \citenamefont {Touboul},\ and\ \citenamefont
  {Zaarcone}}]{lenormand1988}%
  \BibitemOpen
  \bibfield  {author} {\bibinfo {author} {\bibfnamefont {R.}~\bibnamefont
  {Lenormand}}, \bibinfo {author} {\bibfnamefont {E.}~\bibnamefont {Touboul}},\
  and\ \bibinfo {author} {\bibfnamefont {C.}~\bibnamefont {Zaarcone}},\
  }\bibfield  {title} {\bibinfo {title} {Numerical models and experiments on
  immiscible displacement in porous media},\ }\href@noop {} {\bibfield
  {journal} {\bibinfo  {journal} {J. Fluid. Mech.}\ }\textbf {\bibinfo {volume}
  {189}},\ \bibinfo {pages} {165} (\bibinfo {year} {1988})}\BibitemShut
  {NoStop}%
\bibitem [{\citenamefont {L{\o}voll}\ \emph {et~al.}(2004)\citenamefont
  {L{\o}voll}, \citenamefont {M{\`e}heust}, \citenamefont {Toussaint},
  \citenamefont {Schmittbuhl},\ and\ \citenamefont
  {M{\aa}l{\o}y}}]{lovoll2004}%
  \BibitemOpen
  \bibfield  {author} {\bibinfo {author} {\bibfnamefont {G.}~\bibnamefont
  {L{\o}voll}}, \bibinfo {author} {\bibfnamefont {Y.}~\bibnamefont
  {M{\`e}heust}}, \bibinfo {author} {\bibfnamefont {R.}~\bibnamefont
  {Toussaint}}, \bibinfo {author} {\bibfnamefont {J.}~\bibnamefont
  {Schmittbuhl}},\ and\ \bibinfo {author} {\bibfnamefont {K.~J.}\ \bibnamefont
  {M{\aa}l{\o}y}},\ }\bibfield  {title} {\bibinfo {title} {Growth activity
  during fingering in a porous hele shaw cell},\ }\href
  {https://doi.org/10.1103/PhysRevE.70.026301} {\bibfield  {journal} {\bibinfo
  {journal} {Phys. Rev. E}\ }\textbf {\bibinfo {volume} {70}},\ \bibinfo
  {pages} {026301} (\bibinfo {year} {2004})}\BibitemShut {NoStop}%
\bibitem [{\citenamefont {Tallakstad}\ \emph
  {et~al.}(2009{\natexlab{a}})\citenamefont {Tallakstad}, \citenamefont
  {Knudsen}, \citenamefont {Ramstad}, \citenamefont {L{\o}voll}, \citenamefont
  {M\aa{}l\o{}y}, \citenamefont {Toussaint},\ and\ \citenamefont
  {Flekk{\o}y}}]{tallakstad2009}%
  \BibitemOpen
  \bibfield  {author} {\bibinfo {author} {\bibfnamefont {K.~T.}\ \bibnamefont
  {Tallakstad}}, \bibinfo {author} {\bibfnamefont {H.~A.}\ \bibnamefont
  {Knudsen}}, \bibinfo {author} {\bibfnamefont {T.}~\bibnamefont {Ramstad}},
  \bibinfo {author} {\bibfnamefont {G.}~\bibnamefont {L{\o}voll}}, \bibinfo
  {author} {\bibfnamefont {K.~J.}\ \bibnamefont {M\aa{}l\o{}y}}, \bibinfo
  {author} {\bibfnamefont {R.}~\bibnamefont {Toussaint}},\ and\ \bibinfo
  {author} {\bibfnamefont {E.~G.}\ \bibnamefont {Flekk{\o}y}},\ }\bibfield
  {title} {\bibinfo {title} {Steady-state two-phase flow in porous media:
  Statistics and trasnport properties},\ }\href
  {https://doi.org/10.1103/PhysRevLett.102.074502} {\bibfield  {journal}
  {\bibinfo  {journal} {Phys. Rev. Lett.}\ }\textbf {\bibinfo {volume} {102}},\
  \bibinfo {pages} {074502} (\bibinfo {year} {2009}{\natexlab{a}})}\BibitemShut
  {NoStop}%
\bibitem [{\citenamefont {Tallakstad}\ \emph
  {et~al.}(2009{\natexlab{b}})\citenamefont {Tallakstad}, \citenamefont
  {L{\o}voll}, \citenamefont {Knudsen}, \citenamefont {Ramstad}, \citenamefont
  {Flekk{\o}y},\ and\ \citenamefont {M\aa{}l\o{}y}}]{tallakstad2009A}%
  \BibitemOpen
  \bibfield  {author} {\bibinfo {author} {\bibfnamefont {K.~T.}\ \bibnamefont
  {Tallakstad}}, \bibinfo {author} {\bibfnamefont {G.}~\bibnamefont
  {L{\o}voll}}, \bibinfo {author} {\bibfnamefont {H.~A.}\ \bibnamefont
  {Knudsen}}, \bibinfo {author} {\bibfnamefont {T.}~\bibnamefont {Ramstad}},
  \bibinfo {author} {\bibfnamefont {E.~G.}\ \bibnamefont {Flekk{\o}y}},\ and\
  \bibinfo {author} {\bibfnamefont {K.~J.}\ \bibnamefont {M\aa{}l\o{}y}},\
  }\bibfield  {title} {\bibinfo {title} {Steady-state, simultaneous two-phase
  flow in porous media: An experimental study},\ }\href
  {https://doi.org/10.1103/PhysRevE.80.036308} {\bibfield  {journal} {\bibinfo
  {journal} {Phys. Rev. E}\ }\textbf {\bibinfo {volume} {80}},\ \bibinfo
  {pages} {036308} (\bibinfo {year} {2009}{\natexlab{b}})}\BibitemShut
  {NoStop}%
\bibitem [{\citenamefont {Lenormand}\ and\ \citenamefont
  {Zarcone}(1985)}]{lenormand1985}%
  \BibitemOpen
  \bibfield  {author} {\bibinfo {author} {\bibfnamefont {R.}~\bibnamefont
  {Lenormand}}\ and\ \bibinfo {author} {\bibfnamefont {C.}~\bibnamefont
  {Zarcone}},\ }\bibfield  {title} {\bibinfo {title} {Invasion percolation in
  an etched network: Measurment of a fractal dimension},\ }\href
  {https://doi.org/10.1103/PhysRevLett.54.2226} {\bibfield  {journal} {\bibinfo
   {journal} {Phys. Rev. Lett}\ }\textbf {\bibinfo {volume} {54}},\ \bibinfo
  {pages} {2226} (\bibinfo {year} {1985})}\BibitemShut {NoStop}%
\bibitem [{\citenamefont {Furuberg}\ \emph {et~al.}(1988)\citenamefont
  {Furuberg}, \citenamefont {Feder}, \citenamefont {Aharony},\ and\
  \citenamefont {J{\o}ssang}}]{furuberg1988}%
  \BibitemOpen
  \bibfield  {author} {\bibinfo {author} {\bibfnamefont {L.}~\bibnamefont
  {Furuberg}}, \bibinfo {author} {\bibfnamefont {J.}~\bibnamefont {Feder}},
  \bibinfo {author} {\bibfnamefont {A.}~\bibnamefont {Aharony}},\ and\ \bibinfo
  {author} {\bibfnamefont {T.}~\bibnamefont {J{\o}ssang}},\ }\bibfield  {title}
  {\bibinfo {title} {Dynamics of invasion percolation},\ }\href@noop {}
  {\bibfield  {journal} {\bibinfo  {journal} {Phys.\ Rev. Lett.}\ }\textbf
  {\bibinfo {volume} {61}},\ \bibinfo {pages} {2117} (\bibinfo {year}
  {1988})}\BibitemShut {NoStop}%
\bibitem [{\citenamefont {Moura}\ \emph
  {et~al.}(2017{\natexlab{a}})\citenamefont {Moura}, \citenamefont
  {M{\aa}l{\o}y}, \citenamefont {Flekk{\o}y},\ and\ \citenamefont
  {Toussaint}}]{moura2017A}%
  \BibitemOpen
  \bibfield  {author} {\bibinfo {author} {\bibfnamefont {M.}~\bibnamefont
  {Moura}}, \bibinfo {author} {\bibfnamefont {K.~J.}\ \bibnamefont
  {M{\aa}l{\o}y}}, \bibinfo {author} {\bibfnamefont {E.~G.}\ \bibnamefont
  {Flekk{\o}y}},\ and\ \bibinfo {author} {\bibfnamefont {R.}~\bibnamefont
  {Toussaint}},\ }\bibfield  {title} {\bibinfo {title} {Verification of a
  dynamic scaling for the pair correlation function during the slow drainage of
  a porous medium},\ }\href {https://doi.org/10.1103/PhysRevLett.119.154503}
  {\bibfield  {journal} {\bibinfo  {journal} {Phys. Rev. Lett}\ }\textbf
  {\bibinfo {volume} {119}},\ \bibinfo {pages} {154503} (\bibinfo {year}
  {2017}{\natexlab{a}})}\BibitemShut {NoStop}%
\bibitem [{\citenamefont {Xiao}\ \emph {et~al.}(2021)\citenamefont {Xiao},
  \citenamefont {Huang}, \citenamefont {Chen}, \citenamefont {Chen},\ and\
  \citenamefont {Long}}]{xiao2021}%
  \BibitemOpen
  \bibfield  {author} {\bibinfo {author} {\bibfnamefont {B.}~\bibnamefont
  {Xiao}}, \bibinfo {author} {\bibfnamefont {Q.}~\bibnamefont {Huang}},
  \bibinfo {author} {\bibfnamefont {H.}~\bibnamefont {Chen}}, \bibinfo {author}
  {\bibfnamefont {X.}~\bibnamefont {Chen}},\ and\ \bibinfo {author}
  {\bibfnamefont {G.}~\bibnamefont {Long}},\ }\bibfield  {title} {\bibinfo
  {title} {A fractal model for capillary flow through a single tortuous
  capillary with roughened surfaces in fibrous porous media},\ }\href
  {https://doi.org/10.1142/S0218348X21500171} {\bibfield  {journal} {\bibinfo
  {journal} {Fractals}\ }\textbf {\bibinfo {volume} {29}},\ \bibinfo {pages}
  {2150017} (\bibinfo {year} {2021})},\ \Eprint
  {https://arxiv.org/abs/https://doi.org/10.1142/S0218348X21500171}
  {https://doi.org/10.1142/S0218348X21500171} \BibitemShut {NoStop}%
\bibitem [{\citenamefont {Wilkinson}(1984)}]{wilkinson1984}%
  \BibitemOpen
  \bibfield  {author} {\bibinfo {author} {\bibfnamefont {D.}~\bibnamefont
  {Wilkinson}},\ }\bibfield  {title} {\bibinfo {title} {Percolation model of
  immiscible displacement displacement in the presence of buoyancy forces},\
  }\href {https://doi.org/10.1103/PhysRevA.30.520} {\bibfield  {journal}
  {\bibinfo  {journal} {Phys. Rev. A}\ }\textbf {\bibinfo {volume} {30}},\
  \bibinfo {pages} {520} (\bibinfo {year} {1984})}\BibitemShut {NoStop}%
\bibitem [{\citenamefont {Birovljev}\ \emph {et~al.}(1991)\citenamefont
  {Birovljev}, \citenamefont {Furuberg}, \citenamefont {Feder}, \citenamefont
  {J{\o}ssang}, \citenamefont {M{\aa}l{\o}y},\ and\ \citenamefont
  {Aharony}}]{birovljev91}%
  \BibitemOpen
  \bibfield  {author} {\bibinfo {author} {\bibfnamefont {A.}~\bibnamefont
  {Birovljev}}, \bibinfo {author} {\bibfnamefont {L.}~\bibnamefont {Furuberg}},
  \bibinfo {author} {\bibfnamefont {J.}~\bibnamefont {Feder}}, \bibinfo
  {author} {\bibfnamefont {T.}~\bibnamefont {J{\o}ssang}}, \bibinfo {author}
  {\bibfnamefont {K.~J.}\ \bibnamefont {M{\aa}l{\o}y}},\ and\ \bibinfo {author}
  {\bibfnamefont {A.}~\bibnamefont {Aharony}},\ }\bibfield  {title} {\bibinfo
  {title} {Gravity invasion percolation in two dimensions: Experiment and
  simulation},\ }\href@noop {} {\bibfield  {journal} {\bibinfo  {journal}
  {Phys.\ Rev. Lett.}\ }\textbf {\bibinfo {volume} {67}},\ \bibinfo {pages}
  {584} (\bibinfo {year} {1991})}\BibitemShut {NoStop}%
\bibitem [{\citenamefont {Frette}\ \emph {et~al.}(1992)\citenamefont {Frette},
  \citenamefont {Feder}, \citenamefont {J{\o}ssang},\ and\ \citenamefont
  {Meakin}}]{frette1992}%
  \BibitemOpen
  \bibfield  {author} {\bibinfo {author} {\bibfnamefont {V.}~\bibnamefont
  {Frette}}, \bibinfo {author} {\bibfnamefont {J.}~\bibnamefont {Feder}},
  \bibinfo {author} {\bibfnamefont {T.}~\bibnamefont {J{\o}ssang}},\ and\
  \bibinfo {author} {\bibfnamefont {P.}~\bibnamefont {Meakin}},\ }\bibfield
  {title} {\bibinfo {title} {Buoyancy fluid migration in porous media},\ }\href
  {https://doi.org/10.1103/Phys.Rev.Lett.68.3164} {\bibfield  {journal}
  {\bibinfo  {journal} {Phys. Rev. Lett.}\ }\textbf {\bibinfo {volume} {68}},\
  \bibinfo {pages} {3164} (\bibinfo {year} {1992})}\BibitemShut {NoStop}%
\bibitem [{\citenamefont {Wagner}\ \emph {et~al.}(1997)\citenamefont {Wagner},
  \citenamefont {Birovljev}, \citenamefont {Meakin}, \citenamefont {Feder},\
  and\ \citenamefont {J{\o}ssang}}]{wagner1997}%
  \BibitemOpen
  \bibfield  {author} {\bibinfo {author} {\bibfnamefont {G.}~\bibnamefont
  {Wagner}}, \bibinfo {author} {\bibfnamefont {A.}~\bibnamefont {Birovljev}},
  \bibinfo {author} {\bibfnamefont {P.}~\bibnamefont {Meakin}}, \bibinfo
  {author} {\bibfnamefont {J.}~\bibnamefont {Feder}},\ and\ \bibinfo {author}
  {\bibfnamefont {T.}~\bibnamefont {J{\o}ssang}},\ }\bibfield  {title}
  {\bibinfo {title} {Fragmentation and migration of invasion percolation
  clusters:experiments and simulations},\ }\href
  {https://doi.org/10.1103/PhysRevE.55.7015} {\bibfield  {journal} {\bibinfo
  {journal} {Phys.\ Rev. E}\ }\textbf {\bibinfo {volume} {55}},\ \bibinfo
  {pages} {7015} (\bibinfo {year} {1997})}\BibitemShut {NoStop}%
\bibitem [{\citenamefont {Auradou}\ \emph {et~al.}(1999)\citenamefont
  {Auradou}, \citenamefont {M{\aa}l{\o}y}, \citenamefont {Schmittbuhl},
  \citenamefont {Hansen},\ and\ \citenamefont {Bideau}}]{auradou1999}%
  \BibitemOpen
  \bibfield  {author} {\bibinfo {author} {\bibfnamefont {H.}~\bibnamefont
  {Auradou}}, \bibinfo {author} {\bibfnamefont {K.~J.}\ \bibnamefont
  {M{\aa}l{\o}y}}, \bibinfo {author} {\bibfnamefont {J.}~\bibnamefont
  {Schmittbuhl}}, \bibinfo {author} {\bibfnamefont {A.}~\bibnamefont
  {Hansen}},\ and\ \bibinfo {author} {\bibfnamefont {D.}~\bibnamefont
  {Bideau}},\ }\bibfield  {title} {\bibinfo {title} {Competition between
  correlated buoyancy and uncorrelated capillary effects during drainage},\
  }\href@noop {} {\bibfield  {journal} {\bibinfo  {journal} {Phys.\ Rev. E}\
  }\textbf {\bibinfo {volume} {60}},\ \bibinfo {pages} {7224} (\bibinfo {year}
  {1999})}\BibitemShut {NoStop}%
\bibitem [{\citenamefont {Muharrik}\ \emph {et~al.}(2018)\citenamefont
  {Muharrik}, \citenamefont {Suekane},\ and\ \citenamefont
  {Patmonoaji}}]{muharrik2018}%
  \BibitemOpen
  \bibfield  {author} {\bibinfo {author} {\bibfnamefont {M.}~\bibnamefont
  {Muharrik}}, \bibinfo {author} {\bibfnamefont {T.}~\bibnamefont {Suekane}},\
  and\ \bibinfo {author} {\bibfnamefont {A.}~\bibnamefont {Patmonoaji}},\
  }\bibfield  {title} {\bibinfo {title} {Effect of buoyancy on fingering growth
  activity in immiscible two-phase flow displacement},\ }\href
  {https://doi.org/10.1299/jfst.2018jfst0006} {\bibfield  {journal} {\bibinfo
  {journal} {Journal of Fluid Science and Technology}\ }\textbf {\bibinfo
  {volume} {13}},\ \bibinfo {pages} {17} (\bibinfo {year} {2018})}\BibitemShut
  {NoStop}%
\bibitem [{\citenamefont {Zhao}\ \emph {et~al.}(2016)\citenamefont {Zhao},
  \citenamefont {Mac~Minn},\ and\ \citenamefont {Juanes}}]{zhao2016}%
  \BibitemOpen
  \bibfield  {author} {\bibinfo {author} {\bibfnamefont {B.}~\bibnamefont
  {Zhao}}, \bibinfo {author} {\bibfnamefont {C.~W.}\ \bibnamefont {Mac~Minn}},\
  and\ \bibinfo {author} {\bibfnamefont {R.}~\bibnamefont {Juanes}},\
  }\bibfield  {title} {\bibinfo {title} {Wettability control on multiphase flow
  in patterned microfluidics},\ }\href
  {https://doi.org/10.1073/pnas.1603387113} {\bibfield  {journal} {\bibinfo
  {journal} {PNAS}\ }\textbf {\bibinfo {volume} {113}},\ \bibinfo {pages}
  {10251} (\bibinfo {year} {2016})}\BibitemShut {NoStop}%
\bibitem [{\citenamefont {Cottin}\ \emph {et~al.}(2011)\citenamefont {Cottin},
  \citenamefont {Bodiguel},\ and\ \citenamefont {Colin}}]{cottin2011}%
  \BibitemOpen
  \bibfield  {author} {\bibinfo {author} {\bibfnamefont {C.}~\bibnamefont
  {Cottin}}, \bibinfo {author} {\bibfnamefont {H.}~\bibnamefont {Bodiguel}},\
  and\ \bibinfo {author} {\bibfnamefont {A.}~\bibnamefont {Colin}},\ }\bibfield
   {title} {\bibinfo {title} {Influence of wetting conditions on drainage in
  porous media: A microfluidic study},\ }\href
  {https://doi.org/10.1103/Phys.RevE.84.026311} {\bibfield  {journal} {\bibinfo
   {journal} {Phys. Rev. E}\ }\textbf {\bibinfo {volume} {84}},\ \bibinfo
  {pages} {026311} (\bibinfo {year} {2011})}\BibitemShut {NoStop}%
\bibitem [{\citenamefont {Holtzman}\ and\ \citenamefont
  {Segre}(2015)}]{holtzman2015}%
  \BibitemOpen
  \bibfield  {author} {\bibinfo {author} {\bibfnamefont {R.}~\bibnamefont
  {Holtzman}}\ and\ \bibinfo {author} {\bibfnamefont {E.}~\bibnamefont
  {Segre}},\ }\bibfield  {title} {\bibinfo {title} {Wettability stabilizes
  fluid invasion into porous media via nonlocal, cooperative pore filling},\
  }\href {https://doi.org/10.1103/PhysRevLett.115.164501} {\bibfield  {journal}
  {\bibinfo  {journal} {Phys. Rev. Lett.}\ }\textbf {\bibinfo {volume} {115}},\
  \bibinfo {pages} {164501} (\bibinfo {year} {2015})}\BibitemShut {NoStop}%
\bibitem [{\citenamefont {Holtzman}\ \emph {et~al.}(2020)\citenamefont
  {Holtzman}, \citenamefont {Dentz}, \citenamefont {Planet},\ and\
  \citenamefont {Ortin}}]{holtzman2020}%
  \BibitemOpen
  \bibfield  {author} {\bibinfo {author} {\bibfnamefont {R.}~\bibnamefont
  {Holtzman}}, \bibinfo {author} {\bibfnamefont {M.}~\bibnamefont {Dentz}},
  \bibinfo {author} {\bibfnamefont {R.}~\bibnamefont {Planet}},\ and\ \bibinfo
  {author} {\bibfnamefont {J.}~\bibnamefont {Ortin}},\ }\bibfield  {title}
  {\bibinfo {title} {The origin of hysteresis and memory of two-phase flow in
  disordered media},\ }\href {https://doi.org/10.1038/s42005-020-00492-1}
  {\bibfield  {journal} {\bibinfo  {journal} {Communications physics}\ }\textbf
  {\bibinfo {volume} {3}},\ \bibinfo {pages} {222} (\bibinfo {year}
  {2020})}\BibitemShut {NoStop}%
\bibitem [{\citenamefont {de~Gennes}\ \emph {et~al.}(2004)\citenamefont
  {de~Gennes}, \citenamefont {Brochard-Wyart},\ and\ \citenamefont
  {Quéré}}]{gennes2004}%
  \BibitemOpen
  \bibfield  {author} {\bibinfo {author} {\bibfnamefont {P.}~\bibnamefont
  {de~Gennes}}, \bibinfo {author} {\bibfnamefont {F.}~\bibnamefont
  {Brochard-Wyart}},\ and\ \bibinfo {author} {\bibfnamefont {D.}~\bibnamefont
  {Quéré}},\ }\href@noop {} {\emph {\bibinfo {title} {Capillarity and wetting
  phenomena: drops, bubbles, pearls, waves}}}\ (\bibinfo  {publisher}
  {Springer},\ \bibinfo {address} {New York},\ \bibinfo {year}
  {2004})\BibitemShut {NoStop}%
\bibitem [{\citenamefont {Primkulov}\ \emph {et~al.}(2018)\citenamefont
  {Primkulov}, \citenamefont {Talman}, \citenamefont {Khaleghi}, \citenamefont
  {Rangriz~Shokri}, \citenamefont {Chalaturnyk}, \citenamefont {Zhao},
  \citenamefont {MacMinn},\ and\ \citenamefont {Juanes}}]{primkulov2018}%
  \BibitemOpen
  \bibfield  {author} {\bibinfo {author} {\bibfnamefont {B.~K.}\ \bibnamefont
  {Primkulov}}, \bibinfo {author} {\bibfnamefont {S.}~\bibnamefont {Talman}},
  \bibinfo {author} {\bibfnamefont {K.}~\bibnamefont {Khaleghi}}, \bibinfo
  {author} {\bibfnamefont {A.}~\bibnamefont {Rangriz~Shokri}}, \bibinfo
  {author} {\bibfnamefont {R.}~\bibnamefont {Chalaturnyk}}, \bibinfo {author}
  {\bibfnamefont {B.}~\bibnamefont {Zhao}}, \bibinfo {author} {\bibfnamefont
  {C.~W.}\ \bibnamefont {MacMinn}},\ and\ \bibinfo {author} {\bibfnamefont
  {R.}~\bibnamefont {Juanes}},\ }\bibfield  {title} {\bibinfo {title}
  {Quasistatic fluid-fluid displacement in porous media: Invasion-percolation
  through a wetting transition},\ }\href
  {https://doi.org/10.1103/PhysRevFluids.3.104001} {\bibfield  {journal}
  {\bibinfo  {journal} {Phys. Rev. Fluids}\ }\textbf {\bibinfo {volume} {3}},\
  \bibinfo {pages} {104001} (\bibinfo {year} {2018})}\BibitemShut {NoStop}%
\bibitem [{\citenamefont {Primkulov}\ \emph {et~al.}(2019)\citenamefont
  {Primkulov}, \citenamefont {Pahlavan}, \citenamefont {Fu}, \citenamefont
  {Zhao}, \citenamefont {MacMinn},\ and\ \citenamefont
  {Juanes}}]{primkulov2019}%
  \BibitemOpen
  \bibfield  {author} {\bibinfo {author} {\bibfnamefont {B.~K.}\ \bibnamefont
  {Primkulov}}, \bibinfo {author} {\bibfnamefont {A.~A.}\ \bibnamefont
  {Pahlavan}}, \bibinfo {author} {\bibfnamefont {X.}~\bibnamefont {Fu}},
  \bibinfo {author} {\bibfnamefont {B.}~\bibnamefont {Zhao}}, \bibinfo {author}
  {\bibfnamefont {C.~W.}\ \bibnamefont {MacMinn}},\ and\ \bibinfo {author}
  {\bibfnamefont {R.}~\bibnamefont {Juanes}},\ }\bibfield  {title} {\bibinfo
  {title} {Signatures of fluid–fluid displacement in porous media:
  wettability, patterns and pressures},\ }\href
  {https://doi.org/10.1017/jfm.2019.554} {\bibfield  {journal} {\bibinfo
  {journal} {Journal of Fluid Mechanics}\ }\textbf {\bibinfo {volume} {875}},\
  \bibinfo {pages} {R4} (\bibinfo {year} {2019})}\BibitemShut {NoStop}%
\bibitem [{\citenamefont {Rabbani}\ \emph {et~al.}(2018)\citenamefont
  {Rabbani}, \citenamefont {Or}, \citenamefont {Liu}, \citenamefont {Lai},
  \citenamefont {Lu}, \citenamefont {Datta},\ and\ \citenamefont
  {Stone}}]{rabbani2018}%
  \BibitemOpen
  \bibfield  {author} {\bibinfo {author} {\bibfnamefont {H.~S.}\ \bibnamefont
  {Rabbani}}, \bibinfo {author} {\bibfnamefont {D.}~\bibnamefont {Or}},
  \bibinfo {author} {\bibfnamefont {Y.}~\bibnamefont {Liu}}, \bibinfo {author}
  {\bibfnamefont {C.-Y.}\ \bibnamefont {Lai}}, \bibinfo {author} {\bibfnamefont
  {N.~B.}\ \bibnamefont {Lu}}, \bibinfo {author} {\bibfnamefont {S.~S.}\
  \bibnamefont {Datta}},\ and\ \bibinfo {author} {\bibfnamefont {H.~A.}\
  \bibnamefont {Stone}},\ }\bibfield  {title} {\bibinfo {title} {Suppressing
  vicous fingering in structures porous media},\ }\href@noop {} {\bibfield
  {journal} {\bibinfo  {journal} {PNAS}\ }\textbf {\bibinfo {volume} {115}},\
  \bibinfo {pages} {4833} (\bibinfo {year} {2018})}\BibitemShut {NoStop}%
\bibitem [{\citenamefont {Lu}\ \emph {et~al.}(2019)\citenamefont {Lu},
  \citenamefont {Browne},\ and\ \citenamefont {Amchin}}]{lu2019}%
  \BibitemOpen
  \bibfield  {author} {\bibinfo {author} {\bibfnamefont {N.~B.}\ \bibnamefont
  {Lu}}, \bibinfo {author} {\bibfnamefont {C.~A.}\ \bibnamefont {Browne}},\
  and\ \bibinfo {author} {\bibfnamefont {D.~B.}\ \bibnamefont {Amchin}},\
  }\bibfield  {title} {\bibinfo {title} {Controlling capillary fingering using
  pore size gradients in disordered media},\ }\href
  {https://doi.org/10.1103/PhysRevFluids.4.084303} {\bibfield  {journal}
  {\bibinfo  {journal} {Physical Review Fluids}\ }\textbf {\bibinfo {volume}
  {4}},\ \bibinfo {pages} {084303} (\bibinfo {year} {2019})}\BibitemShut
  {NoStop}%
\bibitem [{\citenamefont {Lenormand}(1989)}]{lenormand1989}%
  \BibitemOpen
  \bibfield  {author} {\bibinfo {author} {\bibfnamefont {R.}~\bibnamefont
  {Lenormand}},\ }\bibfield  {title} {\bibinfo {title} {Flow through porous
  media: limits of fractal pattern},\ }\href@noop {} {\bibfield  {journal}
  {\bibinfo  {journal} {Proc. R. Soc. Lond. A}\ }\textbf {\bibinfo {volume}
  {423}},\ \bibinfo {pages} {159} (\bibinfo {year} {1989})}\BibitemShut
  {NoStop}%
\bibitem [{\citenamefont {Payatakes}\ \emph {et~al.}(1980)\citenamefont
  {Payatakes}, \citenamefont {Ng},\ and\ \citenamefont
  {Flumerfelt}}]{payatakes1980}%
  \BibitemOpen
  \bibfield  {author} {\bibinfo {author} {\bibfnamefont {A.~C.}\ \bibnamefont
  {Payatakes}}, \bibinfo {author} {\bibfnamefont {K.~M.}\ \bibnamefont {Ng}},\
  and\ \bibinfo {author} {\bibfnamefont {R.~W.}\ \bibnamefont {Flumerfelt}},\
  }\bibfield  {title} {\bibinfo {title} {Oil ganglion dynamics during
  immiscible displacement: Model formulation},\ }\href
  {https://doi.org/https://doi.org/10.1002/aic.690260315} {\bibfield  {journal}
  {\bibinfo  {journal} {AIChE Journal}\ }\textbf {\bibinfo {volume} {26}},\
  \bibinfo {pages} {430} (\bibinfo {year} {1980})}\BibitemShut {NoStop}%
\bibitem [{\citenamefont {Sandnes}\ \emph {et~al.}(2007)\citenamefont
  {Sandnes}, \citenamefont {Knudsen}, \citenamefont {M{\aa}l{\o}y},\ and\
  \citenamefont {Flekk{\o}y}}]{sandnes2007}%
  \BibitemOpen
  \bibfield  {author} {\bibinfo {author} {\bibfnamefont {B.}~\bibnamefont
  {Sandnes}}, \bibinfo {author} {\bibfnamefont {H.~A.}\ \bibnamefont
  {Knudsen}}, \bibinfo {author} {\bibfnamefont {K.~J.}\ \bibnamefont
  {M{\aa}l{\o}y}},\ and\ \bibinfo {author} {\bibfnamefont {E.~G.}\ \bibnamefont
  {Flekk{\o}y}},\ }\bibfield  {title} {\bibinfo {title} {Labyrinth patterns in
  confined granular-fluid systems},\ }\href
  {https://doi.org/10.1103/PhysRevLett.99.038001} {\bibfield  {journal}
  {\bibinfo  {journal} {Phys. Rev. Lett.}\ }\textbf {\bibinfo {volume} {99}},\
  \bibinfo {pages} {038001} (\bibinfo {year} {2007})}\BibitemShut {NoStop}%
\bibitem [{\citenamefont {Sandnes}\ \emph {et~al.}(2011)\citenamefont
  {Sandnes}, \citenamefont {Flekk{\o}y}, \citenamefont {M{\aa}l{\o}y},\ and\
  \citenamefont {See}}]{sandnes2011}%
  \BibitemOpen
  \bibfield  {author} {\bibinfo {author} {\bibfnamefont {B.}~\bibnamefont
  {Sandnes}}, \bibinfo {author} {\bibfnamefont {E.~G.}\ \bibnamefont
  {Flekk{\o}y}}, \bibinfo {author} {\bibfnamefont {K.~J.}\ \bibnamefont
  {M{\aa}l{\o}y}},\ and\ \bibinfo {author} {\bibfnamefont {H.}~\bibnamefont
  {See}},\ }\bibfield  {title} {\bibinfo {title} {Patterns and flow in
  frictional fluid dynamics},\ }\href {https://doi.org/10.1038/ncomms1289}
  {\bibfield  {journal} {\bibinfo  {journal} {Nat. Commun}\ }\textbf {\bibinfo
  {volume} {2}},\ \bibinfo {pages} {288} (\bibinfo {year} {2011})}\BibitemShut
  {NoStop}%
\bibitem [{\citenamefont {Odier}\ \emph {et~al.}(2017)\citenamefont {Odier},
  \citenamefont {Levach\'e}, \citenamefont {Santanach-Carreras},\ and\
  \citenamefont {Bartolo}}]{odier2017}%
  \BibitemOpen
  \bibfield  {author} {\bibinfo {author} {\bibfnamefont {C.}~\bibnamefont
  {Odier}}, \bibinfo {author} {\bibfnamefont {B.}~\bibnamefont {Levach\'e}},
  \bibinfo {author} {\bibfnamefont {E.}~\bibnamefont {Santanach-Carreras}},\
  and\ \bibinfo {author} {\bibfnamefont {D.}~\bibnamefont {Bartolo}},\
  }\bibfield  {title} {\bibinfo {title} {Forced imbibition in porous media: A
  fourfold scenario},\ }\href {https://doi.org/10.1103/PhysRevLett.119.208005}
  {\bibfield  {journal} {\bibinfo  {journal} {Phys. Rev. Lett.}\ }\textbf
  {\bibinfo {volume} {119}},\ \bibinfo {pages} {208005} (\bibinfo {year}
  {2017})}\BibitemShut {NoStop}%
\bibitem [{\citenamefont {Ferer}\ \emph {et~al.}(1993)\citenamefont {Ferer},
  \citenamefont {Sams}, \citenamefont {R.A.},\ and\ \citenamefont
  {Smith}}]{ferer1993}%
  \BibitemOpen
  \bibfield  {author} {\bibinfo {author} {\bibfnamefont {M.}~\bibnamefont
  {Ferer}}, \bibinfo {author} {\bibfnamefont {W.~N.}\ \bibnamefont {Sams}},
  \bibinfo {author} {\bibfnamefont {G.}~\bibnamefont {R.A.}},\ and\ \bibinfo
  {author} {\bibfnamefont {D.~H.}\ \bibnamefont {Smith}},\ }\bibfield  {title}
  {\bibinfo {title} {Crossover from fractal to compact growth from simulations
  of two-phase flow with finite viscosity ratio in two-dimensional porous
  media},\ }\href@noop {} {\bibfield  {journal} {\bibinfo  {journal} {Phys.
  Rev. E}\ }\textbf {\bibinfo {volume} {47}},\ \bibinfo {pages} {2713}
  (\bibinfo {year} {1993})}\BibitemShut {NoStop}%
\bibitem [{\citenamefont {Ferer}\ \emph {et~al.}(2003)\citenamefont {Ferer},
  \citenamefont {Bromhal},\ and\ \citenamefont {Smith}}]{ferer2003}%
  \BibitemOpen
  \bibfield  {author} {\bibinfo {author} {\bibfnamefont {M.}~\bibnamefont
  {Ferer}}, \bibinfo {author} {\bibfnamefont {G.~S.}\ \bibnamefont {Bromhal}},\
  and\ \bibinfo {author} {\bibfnamefont {D.~H.}\ \bibnamefont {Smith}},\
  }\bibfield  {title} {\bibinfo {title} {Pore-level modeling of drainage:
  Crossover from invasion percolation to compact flow},\ }\href
  {https://doi.org/10.1103/PhysRevE.67.051601} {\bibfield  {journal} {\bibinfo
  {journal} {Phys. Rev. E}\ }\textbf {\bibinfo {volume} {67}},\ \bibinfo
  {pages} {051601} (\bibinfo {year} {2003})}\BibitemShut {NoStop}%
\bibitem [{\citenamefont {Mandelbrot}(1982)}]{mandelbrot1982}%
  \BibitemOpen
  \bibfield  {author} {\bibinfo {author} {\bibfnamefont {B.~B.}\ \bibnamefont
  {Mandelbrot}},\ }\bibfield  {title} {\bibinfo {title} {The fractal geometry
  of nature},\ }\href@noop {} {\bibfield  {journal} {\bibinfo  {journal} {W. H.
  Freeman. San Fransisco}\ } (\bibinfo {year} {1982})}\BibitemShut {NoStop}%
\bibitem [{\citenamefont {Feder}(1988)}]{feder1988}%
  \BibitemOpen
  \bibfield  {author} {\bibinfo {author} {\bibfnamefont {J.}~\bibnamefont
  {Feder}},\ }\bibfield  {title} {\bibinfo {title} {Fractals},\ }\href@noop {}
  {\bibfield  {journal} {\bibinfo  {journal} {Plenum, New York}\ } (\bibinfo
  {year} {1988})}\BibitemShut {NoStop}%
\bibitem [{\citenamefont {Yu}(2008)}]{yu2008}%
  \BibitemOpen
  \bibfield  {author} {\bibinfo {author} {\bibfnamefont {B.}~\bibnamefont
  {Yu}},\ }\bibfield  {title} {\bibinfo {title} {Analysis of flow in fractal
  porous media},\ }\href {https://doi.org/10.1115/1.2955849} {\bibfield
  {journal} {\bibinfo  {journal} {Applied Mechanics Reviews}\ }\textbf
  {\bibinfo {volume} {61}},\ \bibinfo {pages} {050801} (\bibinfo {year}
  {2008})}\BibitemShut {NoStop}%
\bibitem [{\citenamefont {Xiao}\ \emph {et~al.}(2019)\citenamefont {Xiao},
  \citenamefont {Wang}, \citenamefont {Zhang}, \citenamefont {Long},
  \citenamefont {Fan}, \citenamefont {Chen},\ and\ \citenamefont
  {Deng}}]{xiao2019}%
  \BibitemOpen
  \bibfield  {author} {\bibinfo {author} {\bibfnamefont {B.}~\bibnamefont
  {Xiao}}, \bibinfo {author} {\bibfnamefont {W.}~\bibnamefont {Wang}}, \bibinfo
  {author} {\bibfnamefont {X.}~\bibnamefont {Zhang}}, \bibinfo {author}
  {\bibfnamefont {G.}~\bibnamefont {Long}}, \bibinfo {author} {\bibfnamefont
  {J.}~\bibnamefont {Fan}}, \bibinfo {author} {\bibfnamefont {H.}~\bibnamefont
  {Chen}},\ and\ \bibinfo {author} {\bibfnamefont {L.}~\bibnamefont {Deng}},\
  }\bibfield  {title} {\bibinfo {title} {A novel fractal solution for
  permeability and kozeny-carman constant of fibrous porous media made up of
  solid particles and porous fibers},\ }\href
  {https://doi.org/https://doi.org/10.1016/j.powtec.2019.03.028} {\bibfield
  {journal} {\bibinfo  {journal} {Powder Technology}\ }\textbf {\bibinfo
  {volume} {349}},\ \bibinfo {pages} {92} (\bibinfo {year} {2019})}\BibitemShut
  {NoStop}%
\bibitem [{\citenamefont {Liang}\ \emph {et~al.}(2019)\citenamefont {Liang},
  \citenamefont {Fu}, \citenamefont {Xiao}, \citenamefont {Luo},\ and\
  \citenamefont {Wang}}]{liang2019}%
  \BibitemOpen
  \bibfield  {author} {\bibinfo {author} {\bibfnamefont {M.}~\bibnamefont
  {Liang}}, \bibinfo {author} {\bibfnamefont {C.}~\bibnamefont {Fu}}, \bibinfo
  {author} {\bibfnamefont {B.}~\bibnamefont {Xiao}}, \bibinfo {author}
  {\bibfnamefont {L.}~\bibnamefont {Luo}},\ and\ \bibinfo {author}
  {\bibfnamefont {Z.}~\bibnamefont {Wang}},\ }\bibfield  {title} {\bibinfo
  {title} {A fractal study for the effective electrolyte diffusion through
  charged porous media},\ }\href
  {https://doi.org/https://doi.org/10.1016/j.ijheatmasstransfer.2019.03.141}
  {\bibfield  {journal} {\bibinfo  {journal} {International Journal of Heat and
  Mass Transfer}\ }\textbf {\bibinfo {volume} {137}},\ \bibinfo {pages} {365}
  (\bibinfo {year} {2019})}\BibitemShut {NoStop}%
\bibitem [{\citenamefont {Armand}\ \emph {et~al.}(2020)\citenamefont {Armand},
  \citenamefont {Axmann}, \citenamefont {Bresser}, \citenamefont {Copley},
  \citenamefont {Edström}, \citenamefont {Ekberg}, \citenamefont {Guyomard},
  \citenamefont {Lestriez}, \citenamefont {Novák}, \citenamefont
  {Petranikova}, \citenamefont {Porcher}, \citenamefont {Trabesinger},
  \citenamefont {Wohlfahrt-Mehrens},\ and\ \citenamefont {Zhang}}]{armand2020}%
  \BibitemOpen
  \bibfield  {author} {\bibinfo {author} {\bibfnamefont {M.}~\bibnamefont
  {Armand}}, \bibinfo {author} {\bibfnamefont {P.}~\bibnamefont {Axmann}},
  \bibinfo {author} {\bibfnamefont {D.}~\bibnamefont {Bresser}}, \bibinfo
  {author} {\bibfnamefont {M.}~\bibnamefont {Copley}}, \bibinfo {author}
  {\bibfnamefont {K.}~\bibnamefont {Edström}}, \bibinfo {author}
  {\bibfnamefont {C.}~\bibnamefont {Ekberg}}, \bibinfo {author} {\bibfnamefont
  {D.}~\bibnamefont {Guyomard}}, \bibinfo {author} {\bibfnamefont
  {B.}~\bibnamefont {Lestriez}}, \bibinfo {author} {\bibfnamefont
  {P.}~\bibnamefont {Novák}}, \bibinfo {author} {\bibfnamefont
  {M.}~\bibnamefont {Petranikova}}, \bibinfo {author} {\bibfnamefont
  {W.}~\bibnamefont {Porcher}}, \bibinfo {author} {\bibfnamefont
  {S.}~\bibnamefont {Trabesinger}}, \bibinfo {author} {\bibfnamefont
  {M.}~\bibnamefont {Wohlfahrt-Mehrens}},\ and\ \bibinfo {author}
  {\bibfnamefont {H.}~\bibnamefont {Zhang}},\ }\bibfield  {title} {\bibinfo
  {title} {Lithium-ion batteries – current state of the art and anticipated
  developments},\ }\href
  {https://doi.org/https://doi.org/10.1016/j.jpowsour.2020.228708} {\bibfield
  {journal} {\bibinfo  {journal} {Journal of Power Sources}\ }\textbf {\bibinfo
  {volume} {479}},\ \bibinfo {pages} {228708} (\bibinfo {year}
  {2020})}\BibitemShut {NoStop}%
\bibitem [{\citenamefont {Darcy}(1856)}]{darcy1856}%
  \BibitemOpen
  \bibfield  {author} {\bibinfo {author} {\bibfnamefont {H.}~\bibnamefont
  {Darcy}},\ }\bibfield  {title} {\bibinfo {title} {Les fontaines publiques de
  dijon},\ }\href@noop {} {\bibfield  {journal} {\bibinfo  {journal} {Dalmont,
  Paris}\ } (\bibinfo {year} {1856})}\BibitemShut {NoStop}%
\bibitem [{\citenamefont {de~Gennes}\ and\ \citenamefont
  {Guyon}(1978)}]{gennes1978}%
  \BibitemOpen
  \bibfield  {author} {\bibinfo {author} {\bibfnamefont {P.}~\bibnamefont
  {de~Gennes}}\ and\ \bibinfo {author} {\bibfnamefont {E.}~\bibnamefont
  {Guyon}},\ }\bibfield  {title} {\bibinfo {title} {Lois generales pour
  línjections dún fluide dans un milieu poreux aleatoire},\ }\href@noop {}
  {\bibfield  {journal} {\bibinfo  {journal} {J. Mech.}\ }\textbf {\bibinfo
  {volume} {17}},\ \bibinfo {pages} {403} (\bibinfo {year} {1978})}\BibitemShut
  {NoStop}%
\bibitem [{\citenamefont {Chandler}\ \emph {et~al.}(1982)\citenamefont
  {Chandler}, \citenamefont {Koplik}, \citenamefont {Lerman},\ and\
  \citenamefont {Willemsen}}]{chandler1982}%
  \BibitemOpen
  \bibfield  {author} {\bibinfo {author} {\bibfnamefont {R.}~\bibnamefont
  {Chandler}}, \bibinfo {author} {\bibfnamefont {J.}~\bibnamefont {Koplik}},
  \bibinfo {author} {\bibfnamefont {K.}~\bibnamefont {Lerman}},\ and\ \bibinfo
  {author} {\bibfnamefont {D.}~\bibnamefont {Willemsen}},\ }\bibfield  {title}
  {\bibinfo {title} {Capillary displacement and percolation in porous media},\
  }\href@noop {} {\bibfield  {journal} {\bibinfo  {journal} {J. Fluid. Mech.}\
  }\textbf {\bibinfo {volume} {119}},\ \bibinfo {pages} {249} (\bibinfo {year}
  {1982})}\BibitemShut {NoStop}%
\bibitem [{\citenamefont {Wilkinson}\ and\ \citenamefont
  {Willemsen}(1983)}]{wilkinson1983}%
  \BibitemOpen
  \bibfield  {author} {\bibinfo {author} {\bibfnamefont {D.}~\bibnamefont
  {Wilkinson}}\ and\ \bibinfo {author} {\bibfnamefont {J.~F.}\ \bibnamefont
  {Willemsen}},\ }\bibfield  {title} {\bibinfo {title} {Invasion percolation: a
  new form of percolation theory},\ }\href@noop {} {\bibfield  {journal}
  {\bibinfo  {journal} {J. Phys A: Math. Gen.}\ }\textbf {\bibinfo {volume}
  {16}},\ \bibinfo {pages} {3365} (\bibinfo {year} {1983})}\BibitemShut
  {NoStop}%
\bibitem [{\citenamefont {Frette}\ \emph {et~al.}(1997)\citenamefont {Frette},
  \citenamefont {M{\aa}l{\o}y},\ and\ \citenamefont {Schmittbuhl}}]{frette97}%
  \BibitemOpen
  \bibfield  {author} {\bibinfo {author} {\bibfnamefont {O.~I.}\ \bibnamefont
  {Frette}}, \bibinfo {author} {\bibfnamefont {K.~J.}\ \bibnamefont
  {M{\aa}l{\o}y}},\ and\ \bibinfo {author} {\bibfnamefont {J.}~\bibnamefont
  {Schmittbuhl}},\ }\bibfield  {title} {\bibinfo {title} {Immiscible
  displacement of viscosity-matched fluids in two-dimensional porous media},\
  }\href@noop {} {\bibfield  {journal} {\bibinfo  {journal} {Phys. Rev. E}\
  }\textbf {\bibinfo {volume} {57}},\ \bibinfo {pages} {2969} (\bibinfo {year}
  {1997})}\BibitemShut {NoStop}%
\bibitem [{\citenamefont {Aker}\ \emph
  {et~al.}(2000{\natexlab{a}})\citenamefont {Aker}, \citenamefont
  {J\o{}rgen~M\aa{}l\o{}y},\ and\ \citenamefont {Hansen}}]{aker2000C}%
  \BibitemOpen
  \bibfield  {author} {\bibinfo {author} {\bibfnamefont {E.}~\bibnamefont
  {Aker}}, \bibinfo {author} {\bibfnamefont {K.}~\bibnamefont
  {J\o{}rgen~M\aa{}l\o{}y}},\ and\ \bibinfo {author} {\bibfnamefont
  {A.}~\bibnamefont {Hansen}},\ }\bibfield  {title} {\bibinfo {title} {Viscous
  stabilization of 2{D} drainage displacements with trapping},\ }\href
  {https://doi.org/10.1103/PhysRevLett.84.4589} {\bibfield  {journal} {\bibinfo
   {journal} {Phys. Rev. Lett.}\ }\textbf {\bibinfo {volume} {84}},\ \bibinfo
  {pages} {4589} (\bibinfo {year} {2000}{\natexlab{a}})}\BibitemShut {NoStop}%
\bibitem [{\citenamefont {Aker}\ \emph
  {et~al.}(2000{\natexlab{b}})\citenamefont {Aker}, \citenamefont
  {M{\aa}l{\o}y},\ and\ \citenamefont {Hansen}}]{aker2000B}%
  \BibitemOpen
  \bibfield  {author} {\bibinfo {author} {\bibfnamefont {E.}~\bibnamefont
  {Aker}}, \bibinfo {author} {\bibfnamefont {K.~J.}\ \bibnamefont
  {M{\aa}l{\o}y}},\ and\ \bibinfo {author} {\bibfnamefont {A.}~\bibnamefont
  {Hansen}},\ }\bibfield  {title} {\bibinfo {title} {Dynamics of stable viscous
  displacement in porous media,},\ }\href@noop {} {\bibfield  {journal}
  {\bibinfo  {journal} {Phys. Rev. E}\ }\textbf {\bibinfo {volume} {61}},\
  \bibinfo {pages} {2936} (\bibinfo {year} {2000}{\natexlab{b}})}\BibitemShut
  {NoStop}%
\bibitem [{\citenamefont {Meheust}\ \emph {et~al.}(2002)\citenamefont
  {Meheust}, \citenamefont {L{\o}voll}, \citenamefont {M{\aa}l{\o}y},\ and\
  \citenamefont {Schmittbuhl}}]{meheust2002}%
  \BibitemOpen
  \bibfield  {author} {\bibinfo {author} {\bibfnamefont {Y.}~\bibnamefont
  {Meheust}}, \bibinfo {author} {\bibfnamefont {G.}~\bibnamefont {L{\o}voll}},
  \bibinfo {author} {\bibfnamefont {K.~J.}\ \bibnamefont {M{\aa}l{\o}y}},\ and\
  \bibinfo {author} {\bibfnamefont {J.}~\bibnamefont {Schmittbuhl}},\
  }\bibfield  {title} {\bibinfo {title} {Interface scaling in a two-dimensional
  porous medium under combined viscous, gravity, and capillary effects},\
  }\href@noop {} {\bibfield  {journal} {\bibinfo  {journal} {Phys.\ Rev. E}\
  }\textbf {\bibinfo {volume} {66}},\ \bibinfo {pages} {051603} (\bibinfo
  {year} {2002})}\BibitemShut {NoStop}%
\bibitem [{\citenamefont {Haines}(1930)}]{haines1930}%
  \BibitemOpen
  \bibfield  {author} {\bibinfo {author} {\bibfnamefont {W.}~\bibnamefont
  {Haines}},\ }\bibfield  {title} {\bibinfo {title} {Studies in the physical
  properties of soil. the hysteresis effect in capillary properties, and the
  modes of moisture distribution associated therewith},\ }\href
  {https://doi.org/10.1017/S002185960008864X} {\bibfield  {journal} {\bibinfo
  {journal} {J. Agric. Sci.}\ }\textbf {\bibinfo {volume} {20}},\ \bibinfo
  {pages} {97} (\bibinfo {year} {1930})}\BibitemShut {NoStop}%
\bibitem [{\citenamefont {M{\aa}l{\o}y}\ \emph {et~al.}(1992)\citenamefont
  {M{\aa}l{\o}y}, \citenamefont {Furuberg}, \citenamefont {Feder},\ and\
  \citenamefont {J{\o}ssang}}]{maloy92}%
  \BibitemOpen
  \bibfield  {author} {\bibinfo {author} {\bibfnamefont {K.~J.}\ \bibnamefont
  {M{\aa}l{\o}y}}, \bibinfo {author} {\bibfnamefont {L.}~\bibnamefont
  {Furuberg}}, \bibinfo {author} {\bibfnamefont {J.}~\bibnamefont {Feder}},\
  and\ \bibinfo {author} {\bibfnamefont {T.}~\bibnamefont {J{\o}ssang}},\
  }\bibfield  {title} {\bibinfo {title} {Dynamics of slow drainage in porous
  media},\ }\href@noop {} {\bibfield  {journal} {\bibinfo  {journal} {Phys.\
  Rev. Lett.}\ }\textbf {\bibinfo {volume} {68}},\ \bibinfo {pages} {2161}
  (\bibinfo {year} {1992})}\BibitemShut {NoStop}%
\bibitem [{\citenamefont {Furuberg}\ \emph {et~al.}(1996)\citenamefont
  {Furuberg}, \citenamefont {M{\aa}l{\o}y},\ and\ \citenamefont
  {Feder}}]{furuberg96}%
  \BibitemOpen
  \bibfield  {author} {\bibinfo {author} {\bibfnamefont {L.}~\bibnamefont
  {Furuberg}}, \bibinfo {author} {\bibfnamefont {K.~J.}\ \bibnamefont
  {M{\aa}l{\o}y}},\ and\ \bibinfo {author} {\bibfnamefont {J.}~\bibnamefont
  {Feder}},\ }\bibfield  {title} {\bibinfo {title} {Intermittent behaviour in
  slow drainage},\ }\href@noop {} {\bibfield  {journal} {\bibinfo  {journal}
  {Phys.\ Rev. E}\ }\textbf {\bibinfo {volume} {53}},\ \bibinfo {pages} {966}
  (\bibinfo {year} {1996})}\BibitemShut {NoStop}%
\bibitem [{\citenamefont {Moebius}\ and\ \citenamefont
  {Or}(2014{\natexlab{a}})}]{moebius2014}%
  \BibitemOpen
  \bibfield  {author} {\bibinfo {author} {\bibfnamefont {F.}~\bibnamefont
  {Moebius}}\ and\ \bibinfo {author} {\bibfnamefont {D.}~\bibnamefont {Or}},\
  }\bibfield  {title} {\bibinfo {title} {Pore scale dynamics underlying the
  motion of drainage fronts in porous media},\ }\href
  {https://doi.org/10.1002/2014WR015916} {\bibfield  {journal} {\bibinfo
  {journal} {Water Resour. Res.}\ }\textbf {\bibinfo {volume} {50}},\ \bibinfo
  {pages} {8441} (\bibinfo {year} {2014}{\natexlab{a}})}\BibitemShut {NoStop}%
\bibitem [{\citenamefont {Moebius}\ and\ \citenamefont
  {Or}(2014{\natexlab{b}})}]{moebius2014a}%
  \BibitemOpen
  \bibfield  {author} {\bibinfo {author} {\bibfnamefont {F.}~\bibnamefont
  {Moebius}}\ and\ \bibinfo {author} {\bibfnamefont {D.}~\bibnamefont {Or}},\
  }\bibfield  {title} {\bibinfo {title} {Inertial forces affect fluid front
  displacement dynamics in a pore-throat network model},\ }\href
  {https://doi.org/10.1103/PhysRevE.90.023019} {\bibfield  {journal} {\bibinfo
  {journal} {Phys. Rev. E,}\ }\textbf {\bibinfo {volume} {90}},\ \bibinfo
  {pages} {023019} (\bibinfo {year} {2014}{\natexlab{b}})}\BibitemShut
  {NoStop}%
\bibitem [{\citenamefont {Moebius}\ \emph {et~al.}(2012)\citenamefont
  {Moebius}, \citenamefont {Canone},\ and\ \citenamefont {Or}}]{moebius2012}%
  \BibitemOpen
  \bibfield  {author} {\bibinfo {author} {\bibfnamefont {F.}~\bibnamefont
  {Moebius}}, \bibinfo {author} {\bibfnamefont {D.}~\bibnamefont {Canone}},\
  and\ \bibinfo {author} {\bibfnamefont {D.}~\bibnamefont {Or}},\ }\bibfield
  {title} {\bibinfo {title} {Characteristics of acoustic emissions induced by
  fluid front displacement in porous media},\ }\href
  {https://doi.org/10.1029/2012WR012525} {\bibfield  {journal} {\bibinfo
  {journal} {Water Resour. Res.}\ }\textbf {\bibinfo {volume} {48}},\ \bibinfo
  {pages} {W11507} (\bibinfo {year} {2012})}\BibitemShut {NoStop}%
\bibitem [{\citenamefont {Berg}\ and\ \citenamefont {et~al.}(2013)}]{berg2013}%
  \BibitemOpen
  \bibfield  {author} {\bibinfo {author} {\bibnamefont {Berg}}\ and\ \bibinfo
  {author} {\bibnamefont {et~al.}},\ }\bibfield  {title} {\bibinfo {title}
  {Real-time 3d imaging of haines jumps in porous media flow},\ }\href
  {https://doi.org/10.1073/pnas.1221373110} {\bibfield  {journal} {\bibinfo
  {journal} {PNAS}\ }\textbf {\bibinfo {volume} {110}},\ \bibinfo {pages}
  {3755} (\bibinfo {year} {2013})}\BibitemShut {NoStop}%
\bibitem [{\citenamefont {Bultreys}\ \emph {et~al.}(2015)\citenamefont
  {Bultreys}, \citenamefont {Boone}, \citenamefont {Boone}, \citenamefont
  {Schryver}, \citenamefont {Masschaele}, \citenamefont {Loo}, \citenamefont
  {Hoorebeke},\ and\ \citenamefont {Cnudde}}]{bultreys2015}%
  \BibitemOpen
  \bibfield  {author} {\bibinfo {author} {\bibfnamefont {T.}~\bibnamefont
  {Bultreys}}, \bibinfo {author} {\bibfnamefont {M.~A.}\ \bibnamefont {Boone}},
  \bibinfo {author} {\bibfnamefont {M.~N.}\ \bibnamefont {Boone}}, \bibinfo
  {author} {\bibfnamefont {T.~D.}\ \bibnamefont {Schryver}}, \bibinfo {author}
  {\bibfnamefont {B.}~\bibnamefont {Masschaele}}, \bibinfo {author}
  {\bibfnamefont {D.~V.}\ \bibnamefont {Loo}}, \bibinfo {author} {\bibfnamefont
  {L.~V.}\ \bibnamefont {Hoorebeke}},\ and\ \bibinfo {author} {\bibfnamefont
  {V.}~\bibnamefont {Cnudde}},\ }\bibfield  {title} {\bibinfo {title}
  {Real-time visualization of haines jumps in sandstone with laboratory-based
  microcomputed tomography},\ }\href {https://doi.org/10.1002/2015WR017502}
  {\bibfield  {journal} {\bibinfo  {journal} {Water Resour. Res.}\ }\textbf
  {\bibinfo {volume} {51}},\ \bibinfo {pages} {8668} (\bibinfo {year}
  {2015})}\BibitemShut {NoStop}%
\bibitem [{\citenamefont {Zacharoudiou}\ \emph {et~al.}(2018)\citenamefont
  {Zacharoudiou}, \citenamefont {Boek},\ and\ \citenamefont
  {Crawshaw}}]{zacharoudiou2018}%
  \BibitemOpen
  \bibfield  {author} {\bibinfo {author} {\bibfnamefont {I.}~\bibnamefont
  {Zacharoudiou}}, \bibinfo {author} {\bibfnamefont {E.~S.}\ \bibnamefont
  {Boek}},\ and\ \bibinfo {author} {\bibfnamefont {J.}~\bibnamefont
  {Crawshaw}},\ }\bibfield  {title} {\bibinfo {title} {The impact of drainage
  displacement patterns and haines jumps on co2 storage efficency},\ }\href
  {https://doi.org/10.1038/s41598-018-33502-y} {\bibfield  {journal} {\bibinfo
  {journal} {Scientific Reports}\ }\textbf {\bibinfo {volume} {8}},\ \bibinfo
  {pages} {15561} (\bibinfo {year} {2018})}\BibitemShut {NoStop}%
\bibitem [{\citenamefont {Moura}\ \emph
  {et~al.}(2017{\natexlab{b}})\citenamefont {Moura}, \citenamefont
  {M{\aa}l{\o}y},\ and\ \citenamefont {Toussaint}}]{moura2017}%
  \BibitemOpen
  \bibfield  {author} {\bibinfo {author} {\bibfnamefont {M.}~\bibnamefont
  {Moura}}, \bibinfo {author} {\bibfnamefont {K.~J.}\ \bibnamefont
  {M{\aa}l{\o}y}},\ and\ \bibinfo {author} {\bibfnamefont {R.}~\bibnamefont
  {Toussaint}},\ }\bibfield  {title} {\bibinfo {title} {Critical behaviour in
  porous media flow},\ }\href@noop {} {\bibfield  {journal} {\bibinfo
  {journal} {Europhys. Lett.}\ }\textbf {\bibinfo {volume} {118}},\ \bibinfo
  {pages} {14004} (\bibinfo {year} {2017}{\natexlab{b}})}\BibitemShut {NoStop}%
\bibitem [{\citenamefont {Berg}\ \emph {et~al.}(2020)\citenamefont {Berg},
  \citenamefont {Slotte},\ and\ \citenamefont {Khanamiri}}]{berg2020}%
  \BibitemOpen
  \bibfield  {author} {\bibinfo {author} {\bibfnamefont {C.~F.}\ \bibnamefont
  {Berg}}, \bibinfo {author} {\bibfnamefont {P.~A.}\ \bibnamefont {Slotte}},\
  and\ \bibinfo {author} {\bibfnamefont {H.~H.}\ \bibnamefont {Khanamiri}},\
  }\bibfield  {title} {\bibinfo {title} {Geometrical derived efficiency of slow
  immiscible displacement in porous media},\ }\href
  {https://doi.org/10.1103/PhysRevE.102.033113} {\bibfield  {journal} {\bibinfo
   {journal} {Phys.\ Rev. E}\ }\textbf {\bibinfo {volume} {102}},\ \bibinfo
  {pages} {033113} (\bibinfo {year} {2020})}\BibitemShut {NoStop}%
\bibitem [{\citenamefont {Hele-Shaw}(1898)}]{heleshaw1898}%
  \BibitemOpen
  \bibfield  {author} {\bibinfo {author} {\bibfnamefont {H.~S.}\ \bibnamefont
  {Hele-Shaw}},\ }\bibfield  {title} {\bibinfo {title} {The flow of water},\
  }\href@noop {} {\bibfield  {journal} {\bibinfo  {journal} {Nature}\ }\textbf
  {\bibinfo {volume} {58}},\ \bibinfo {pages} {34} (\bibinfo {year}
  {1898})}\BibitemShut {NoStop}%
\bibitem [{\citenamefont {Moura}\ \emph {et~al.}(2015)\citenamefont {Moura},
  \citenamefont {Florentino}, \citenamefont {M{\aa}l{\o}y}, \citenamefont
  {Sch{\"a}fer},\ and\ \citenamefont {Toussaint}}]{moura2015}%
  \BibitemOpen
  \bibfield  {author} {\bibinfo {author} {\bibfnamefont {M.}~\bibnamefont
  {Moura}}, \bibinfo {author} {\bibfnamefont {E.}~\bibnamefont {Florentino}},
  \bibinfo {author} {\bibfnamefont {K.}~\bibnamefont {M{\aa}l{\o}y}}, \bibinfo
  {author} {\bibfnamefont {G.}~\bibnamefont {Sch{\"a}fer}},\ and\ \bibinfo
  {author} {\bibfnamefont {R.}~\bibnamefont {Toussaint}},\ }\bibfield  {title}
  {\bibinfo {title} {Impact of sample geometry on the measurement of
  pressure-saturation curves: Experiments and simulations},\ }\bibfield
  {journal} {\bibinfo  {journal} {Water Resour. Res.}\ }\textbf {\bibinfo
  {volume} {51}},\ \href {https://doi.org/doi:10.1002/2015WR017196}
  {doi:10.1002/2015WR017196} (\bibinfo {year} {2015})\BibitemShut {NoStop}%
\bibitem [{\citenamefont {Hinrichsen}\ \emph {et~al.}(1986)\citenamefont
  {Hinrichsen}, \citenamefont {Feder},\ and\ \citenamefont
  {J{\o}ssang}}]{hinrichsen1986}%
  \BibitemOpen
  \bibfield  {author} {\bibinfo {author} {\bibfnamefont {E.~L.}\ \bibnamefont
  {Hinrichsen}}, \bibinfo {author} {\bibfnamefont {J.}~\bibnamefont {Feder}},\
  and\ \bibinfo {author} {\bibfnamefont {T.}~\bibnamefont {J{\o}ssang}},\
  }\bibfield  {title} {\bibinfo {title} {Geometry of random sequential
  adsorption},\ }\href@noop {} {\bibfield  {journal} {\bibinfo  {journal}
  {Jornal of Statistical Physics}\ }\textbf {\bibinfo {volume} {44}},\ \bibinfo
  {pages} {793} (\bibinfo {year} {1986})}\BibitemShut {NoStop}%
\bibitem [{\citenamefont {Stauffer}(1994)}]{stauffer1994}%
  \BibitemOpen
  \bibfield  {author} {\bibinfo {author} {\bibfnamefont {D.}~\bibnamefont
  {Stauffer}},\ }\href@noop {} {\emph {\bibinfo {title} {Introduction to
  percolation theory}}}\ (\bibinfo  {publisher} {Taylor \& Francis},\ \bibinfo
  {address} {London Bristol, PA},\ \bibinfo {year} {1994})\BibitemShut
  {NoStop}%
\bibitem [{\citenamefont {Martys}\ \emph {et~al.}(1991)\citenamefont {Martys},
  \citenamefont {Robbins},\ and\ \citenamefont {Cieplak}}]{martys1991}%
  \BibitemOpen
  \bibfield  {author} {\bibinfo {author} {\bibfnamefont {N.}~\bibnamefont
  {Martys}}, \bibinfo {author} {\bibfnamefont {M.}~\bibnamefont {Robbins}},\
  and\ \bibinfo {author} {\bibfnamefont {M.}~\bibnamefont {Cieplak}},\
  }\bibfield  {title} {\bibinfo {title} {Scaling relations for interface motion
  through disordered media: Application to two-dimensional fluid invasion},\
  }\href@noop {} {\bibfield  {journal} {\bibinfo  {journal} {Phys.\ Rev. B}\
  }\textbf {\bibinfo {volume} {44}},\ \bibinfo {pages} {12294} (\bibinfo {year}
  {1991})}\BibitemShut {NoStop}%
\bibitem [{\citenamefont {Roux}\ and\ \citenamefont {Guyon}(1989)}]{roux1989}%
  \BibitemOpen
  \bibfield  {author} {\bibinfo {author} {\bibfnamefont {S.}~\bibnamefont
  {Roux}}\ and\ \bibinfo {author} {\bibfnamefont {E.}~\bibnamefont {Guyon}},\
  }\bibfield  {title} {\bibinfo {title} {Temporal development of invasion
  percolation},\ }\href {https://doi.org/10.1088/0305-4470/22/17/034}
  {\bibfield  {journal} {\bibinfo  {journal} {J. Phys. A}\ }\textbf {\bibinfo
  {volume} {22}},\ \bibinfo {pages} {3693} (\bibinfo {year}
  {1989})}\BibitemShut {NoStop}%
\bibitem [{\citenamefont {Sapoval}\ \emph {et~al.}(1985)\citenamefont
  {Sapoval}, \citenamefont {Rosso},\ and\ \citenamefont {Gouyet}}]{sapova1985}%
  \BibitemOpen
  \bibfield  {author} {\bibinfo {author} {\bibfnamefont {E.}~\bibnamefont
  {Sapoval}}, \bibinfo {author} {\bibfnamefont {M.}~\bibnamefont {Rosso}},\
  and\ \bibinfo {author} {\bibfnamefont {J.}~\bibnamefont {Gouyet}},\
  }\bibfield  {title} {\bibinfo {title} {The fractal nature of a diffusion
  front and the relation to percolation},\ }\href@noop {} {\bibfield  {journal}
  {\bibinfo  {journal} {J. Phys (France) Lett.}\ }\textbf {\bibinfo {volume}
  {46}},\ \bibinfo {pages} {L149} (\bibinfo {year} {1985})}\BibitemShut
  {NoStop}%
\bibitem [{\citenamefont {Gouyet}\ \emph {et~al.}(1988)\citenamefont {Gouyet},
  \citenamefont {Rosso},\ and\ \citenamefont {Sapoval}}]{gouyet1988}%
  \BibitemOpen
  \bibfield  {author} {\bibinfo {author} {\bibfnamefont {J.}~\bibnamefont
  {Gouyet}}, \bibinfo {author} {\bibfnamefont {M.}~\bibnamefont {Rosso}},\ and\
  \bibinfo {author} {\bibfnamefont {B.}~\bibnamefont {Sapoval}},\ }\bibfield
  {title} {\bibinfo {title} {Fractal structure of diffusion and invasion fronts
  in three-dimensional lattices through the gradient percolation approach},\
  }\href@noop {} {\bibfield  {journal} {\bibinfo  {journal} {Phys. Rev. B}\
  }\textbf {\bibinfo {volume} {37}},\ \bibinfo {pages} {1832} (\bibinfo {year}
  {1988})}\BibitemShut {NoStop}%
\bibitem [{\citenamefont {Ayaz}\ \emph {et~al.}(2020)\citenamefont {Ayaz},
  \citenamefont {Toussaint}, \citenamefont {Sch{\"a}fer},\ and\ \citenamefont
  {M{\aa}l{\o}y}}]{ayaz2020}%
  \BibitemOpen
  \bibfield  {author} {\bibinfo {author} {\bibfnamefont {M.}~\bibnamefont
  {Ayaz}}, \bibinfo {author} {\bibfnamefont {R.}~\bibnamefont {Toussaint}},
  \bibinfo {author} {\bibfnamefont {G.}~\bibnamefont {Sch{\"a}fer}},\ and\
  \bibinfo {author} {\bibfnamefont {K.~J.}\ \bibnamefont {M{\aa}l{\o}y}},\
  }\bibfield  {title} {\bibinfo {title} {Gravitational and finite-size effects
  on pressure saturation curves during drainage},\ }\bibfield  {journal}
  {\bibinfo  {journal} {Water Resour. Res.}\ }\textbf {\bibinfo {volume}
  {56}},\ \href {https://doi.org/e2019WR026279} {e2019WR026279} (\bibinfo
  {year} {2020})\BibitemShut {NoStop}%
\end{thebibliography}
\end{document}